\documentclass[12pt]{article}
\RequirePackage{natbib}
\usepackage{graphicx}
\usepackage{amssymb,amsfonts}
\RequirePackage[colorlinks,citecolor=blue,urlcolor=blue]{hyperref}
\hypersetup{colorlinks = false}
\usepackage{fullpage}
\usepackage{hyperref}
\usepackage{lineno}
\usepackage{etoolbox}
\usepackage{subcaption}
\usepackage{rotating}
\usepackage{float}
\usepackage{color}
\usepackage{xr}
\usepackage{booktabs}
\usepackage{tabulary}
\usepackage[]{algorithm2e}
\restylefloat{table}
\graphicspath{ {img/} }

\newcommand{\beginsupplement}{%
        \setcounter{table}{0}
        \renewcommand{\thetable}{S\arabic{table}}%
        \setcounter{figure}{0}
        \renewcommand{\thefigure}{S\arabic{figure}}%
        \setcounter{section}{0}
        \renewcommand{\thesection}{S\arabic{section}}%
}

\begin{document}

\title{SNIP: An Adaptation of Sorted Neighborhood Methods for Deduplicating Pedigree Data}
\author{Theodore Huang$^{1,2,*}$, Matthew Ploenzke, Danielle Braun$^{1,2}$}
\date{}

\maketitle

\begin{center}
$^{1}$Department of Biostatistics, Harvard T.H. Chan School of Public Health \\
$^{2}$Department of Data Science, Dana-Farber Cancer Institute \\
$^*$Corresponding author: thuang@ds.dfci.harvard.edu
\end{center}

\begin{abstract}
Pedigree data contain family history information that is used to analyze hereditary diseases. These clinical data sets may contain duplicate records due to the same family visiting a clinic multiple times or a clinician entering multiple versions of the family for testing purposes. Inferences drawn from the data or using them for training or validation without removing the duplicates could lead to invalid conclusions, and hence identifying the duplicates is essential. Since family structures can be complex, existing deduplication algorithms cannot be applied directly. We first motivate the importance of deduplication by examining the impact of pedigree duplicates on the training and validation of a familial risk prediction model. We then introduce an unsupervised algorithm, which we call SNIP (\textbf{S}orted \textbf{N}e\textbf{I}ghborhood for \textbf{P}edigrees), that builds on the sorted neighborhood method to efficiently find and classify pairwise comparisons by leveraging the inherent hierarchical nature of the pedigrees. We conduct a simulation study to assess the performance of the algorithm and find parameter configurations where the algorithm is able to accurately detect the duplicates. We then apply the method to data from the Risk Service, which includes over 300,000 pedigrees at high risk of hereditary cancers, and uncover large clusters of potential duplicate families. After removing 104,520 pedigrees (33\% of original data), the resulting Risk Service dataset can now be used for future analysis, training, and validation.  The algorithm is available as an R package \url{snipR} available at \url{https://github.com/bayesmendel/snipR}.
\end{abstract}

\section{Introduction}
\label{sec:intro}

Pedigree data are often collected to study and quantify genetic mechanisms for hereditary diseases. Collecting pedigrees, especially extended pedigrees that span several generations, can be more useful and efficient than collecting data on unrelated individuals, allowing for studying linkage, heritability, and shared environmental effects \citep{williams2006collection}. We focus on pedigree data which include many families, where each family is assigned a family-level ID, and each individual in each family is assigned an individual-level ID. The IDs of the mother and father of each individual are also provided, so that the exact structure of the pedigree can be determined. In addition to the pedigree structure, information on phenotypes, and in some settings genotypes, demographics, and other information, are collected for each individual. An example of a simple pedigree data set containing two families, collected in the context of hereditary breast and ovarian cancer, is provided in Table~\ref{tab:example}. Plots of the two pedigrees are provided in Supplementary Figure~\ref{sfig:example_plot}.

Pedigree data can sometimes include duplicate families. This is the case in the Risk Service data, which motivated this work and the data application in Section~\ref{sec:dataapp}. The Risk Service is a web interface that allows clinicians to input pedigree information and receive risk estimates for genetic susceptibility to hereditary cancers \citep{risk_service}. It provides risk estimates from the following models: the BayesMendel models BRCAPRO, MMRPRO, PANCPRO, MELAPRO \citep{bayesmendel}; the IBIS Breast Cancer Risk Evaluation Tool \citep{tyrer2004breast}; and the Colorectal Cancer Risk Assessment Tool (CCRAT) \citep{freedman2009colorectal}. Each time a clinician uses the software and enters pedigree data, the information is stored in a database on the server. However, no identifying information is stored and hence we cannot determine if the same family has been stored previously in the database. Thus, duplicate pedigrees can occur if, for example, a proband (the individual being counseled from whom we obtain the family history) visits the clinic multiple times, or if a clinician enters multiple versions of the pedigree to assess risk under varying family history scenarios.  The presence of these duplicates can bias analyses, and hence developing algorithms to classify the duplicates is important for valid statistical inference.

The literature refers to algorithms of this sort as data deduplication, duplicate detection, entity matching, and record linkage \citep{survey_2007,comparison_2010,survey_2016,calado2010overview}.
Several methods have been proposed to detect duplicates in pedigree data \citep{pixton2005mal4,pixton2006using,ivie2007metric}, but these are supervised methods requiring training data.
It is thus our goal to expand upon existing deduplication algorithms to incorporate the unique pedigree structures while using unsupervised learning frameworks given our lack of a labeled training set with known duplicates.

In this paper, we adapt the sorted neighborhood method introduced by \cite{hernandez1995merge} to pedigree data. \cite{hernandez1995merge} use a two-phase process of clustering the data into blocks and then sorting within blocks to identify potential duplicates. We apply the algorithm to pedigrees by first splitting the families into predefined family relationship types (probands, mothers, fathers, etc.). Then we split each relationship type into blocks based on the blocking variables, and within each block we sort the families to identify pairs of potential duplicate families. Finally, we classify as duplicates the pairs which were found in a sufficient number of relationship types. See Figure~\ref{fig:flowchart}. The schema is similar to aggregation methods proposed by \citep{mapreduce} and \citep{metablocking}. Since we apply the sorted-neighborhood method to pedigree data, we call our extension the SNIP (\textbf{S}orted \textbf{N}e\textbf{I}ghborhood for \textbf{P}edigrees) algorithm. The algorithm is available as an R package \url{snipR} available at \url{https://github.com/bayesmendel/snipR}, and the simulation code and results are available at \url{https://github.com/theohuang/snipR-Simulations}.

The rest of the paper is structured as follows. Section~\ref{sec:motiv} motivates the importance of deduplication in pedigree data by showing the differences in performance when training and validating prediction models for hereditary cancers in data sets with and without duplicates. Section~\ref{sec:methods} explains the algorithm in detail, Section~\ref{sec:simulations} illustrates the algorithm in a simulated pedigree data set, and Section~\ref{sec:dataapp} applies the algorithm in pedigree data collected through the Risk Service. Finally, the paper concludes with a brief discussion in Section~\ref{sec:discussion}.

\begin{table}
    \centering
    \begin{tiny}
\begin{tabulary}{\textwidth}{RRRRRRRRRRRR}
  \toprule
FamID & ID & MotherID & FatherID & isProband & Sex & CurAge & isAffBC & isAffOC & AgeBC & AgeOC & isDead \\ 
  \midrule
1 & 1 & 5 & 7 & 0 & 0 & NA & 1 & 0 & 65 & NA & 0 \\ 
  1 & 2 & 6 & 8 & 0 & 1 & 75 & 0 & 0 & NA & NA & 1 \\ 
  1 & 3 & 1 & 2 & 1 & 0 & 54 & 0 & 1 & NA & NA & 0 \\ 
  1 & 4 & 1 & 2 & 0 & 1 & 57 & 0 & 0 & NA & NA & 0 \\ 
  1 & 5 & NA & NA & 0 & 0 & 86 & 0 & 0 & NA & NA & 1 \\ 
  1 & 6 & NA & NA & 0 & 0 & 91 & 0 & 0 & NA & NA & 1 \\ 
  1 & 7 & NA & NA & 0 & 1 & 77 & 0 & 0 & NA & NA & 1 \\ 
  1 & 8 & NA & NA & 0 & 1 & 64 & 1 & 0 & 62 & NA & 1 \\
  \midrule[0.01pt]
  2 & 1 & 5 & 7 & 0 & 0 & 84 & 0 & 1 & NA & NA & 0 \\ 
  2 & 2 & 6 & 8 & 0 & 1 & 80 & 0 & 0 & NA & NA & 1 \\ 
  2 & 3 & 1 & 2 & 1 & 0 & 61 & 0 & 0 & NA & NA & 0 \\ 
  2 & 4 & 1 & 2 & 0 & 0 & 58 & 0 & 0 & NA & NA & 0 \\ 
  2 & 5 & NA & NA & 0 & 0 & 52 & 0 & 0 & NA & NA & 1 \\ 
  2 & 6 & NA & NA & 0 & 0 & 83 & 1 & 0 & 61 & NA & 1 \\ 
  2 & 7 & NA & NA & 0 & 1 & NA & 0 & 0 & NA & NA & 1 \\ 
  2 & 8 & NA & NA & 0 & 1 & 71 & 0 & 0 & NA & NA & 1 \\ 
  2 & 9 & NA & NA & 0 & 1 & 64 & 0 & 0 & NA & NA & 0 \\ 
  2 & 10 & 3 & 9 & 0 & 1 & 35 & 0 & 0 & NA & NA & 0 \\ 
  2 & 11 & 3 & 9 & 0 & 0 & 31 & 0 & 0 & NA & NA & 0 \\
  \bottomrule
\end{tabulary}
\end{tiny}
    \caption{Example of a pedigree data set with two families, collected for studying hereditary breast and ovarian cancer. FamID is the ID variable denoting the family. ID is the individual-level ID variable within the family, and MotherID and FatherID provide the IDs of an individual's mother and father (NA for founders of the pedigree), respectively. isProband indicates the proband (the individual being counseled from whom we obtain the family history)  of the family. Sex is 1 for males and 0 for females. CurAge is either the current age (if isDead = 0) or the death age (if isDead = 1). isAffBC and isAffOC are indicators of the individual having developed breast and ovarian cancer, respectively, with AgeBC and AgeOC the corresponding ages of onset (NA if no cancer). The NA's in MotherID and FatherID are expected for the founders, and the NA's in AgeBC and AgeOC are expected except in the case where the individual has the corresponding cancer, in which case it indicates missing information. Supplementary Figure~\ref{sfig:example_plot} shows a plot of this pedigree.}
    \label{tab:example}
\end{table}

\section{Motivation} \label{sec:motiv}

Studies have been conducted showing the impact of duplicates in data analysis, such as \cite{waldron2016doppelganger}. To motivate the impact of deduplication in pedigree data analysis, we conduct a simulation study where we assess the performance of training and validating prediction models on pedigree data with and without duplicates. We can hence quantify the importance of removing duplicates from a data set when training prediction models and evaluating their performance. We explore the following situations: (1) Monte Carlo cross-validation on a data set with duplicates, where we train on a random sample and validate on the other; (2) train the model on a data set with duplicates and validate on a data set without duplicates; and (3) train the model on a data set without duplicates and validate on a data set with duplicates.

Pedigree data are often used as inputs for models predicting an individual's risk of carrying cancer susceptibility gene mutations. We thus focus this section on training logistic regression models to predict an individual's probability of carrying a mutation in at least one of \textit{BRCA1} and \textit{BRCA2}, two cancer susceptibility genes for hereditary breast and ovarian cancer.

\subsection{Data Generation}
\label{subsec:motiv_gen}

We first generate the pedigree data set with duplicates. Generating pedigree data can be achieved in a variety of ways, but we choose to explicitly use Mendelian laws of inheritance since we are illustrating the importance of deduplication in the context of predicting an individual's mutation carrier status. We start by generating \textit{BRCA1} and \textit{BRCA2} carrier statuses. We generate two separate data sets using two levels of allele frequencies: 0.01 and 0.1. To do so, we first generate mutation statuses for the founders of the pedigree based on the specified allele frequency, and then use Mendelian laws of inheritance to generate mutation statuses of the remaining relatives. We then generate breast and ovarian cancer statuses and ages of onset based on the mutation statuses, using literature estimates of age-specific cancer risks conditional on genotypes (penetrances). We use the \texttt{PedUtils} R package (\url{https://github.com/bayesmendel/PedUtils}) for the data generation. The literature estimates of cancer penetrances are the ones used in BRCAPRO \citep{bayesmendel}, a Mendelian risk model for hereditary cancer syndromes.

We generate $n=10,000$ families, each with a proband, siblings, father and mother, paternal and maternal grandparents, paternal and maternal aunts and uncles, children, and nieces and nephews. The number of brothers, sisters, paternal and maternal aunts and uncles, children, and children of each sibling is randomly generated from $\{0, 1, 2, 3\}$.

Each data set, representing the two allele frequency scenarios, consists initially of 10,000 families, without duplicates. For each scenario we then generate a database of 10,000 duplicate families from the original data set, where we sample with replacement and each duplicate represents one copy of a randomly selected family from the original data set. We use this database to draw duplicates under varying levels of duplication. We consider two parameters impacting this level of duplication: (1) the proportion of the original data that consists of duplicates, and (2) the proportion of duplicates that have an extensive family history of breast and ovarian cancer. Here we define ``extensive" as having a proportion of relatives with breast or ovarian cancer that is greater than the 90th percentile in the database of duplicates. For each configuration of these two parameters, we generate an augmented duplicated data set with duplicates. If the proportion of the original data that consists of duplicates is $p_d$ and the proportion of duplicates with extensive family history is $p_{h}$, then we add $\lfloor p_{h}p_d n \rfloor$ families from the database of duplicates with extensive family history and $\lfloor p_d n \rfloor - \lfloor p_{h}p_d n \rfloor$ families from the data set without extensive family history. Thus, in total the data set for this parameter configuration consists of 10,000 original families along with $\lfloor 10,000p_d \rfloor$ duplicates.

In real data, duplicate entries may not be exact copies of the original entry but may contain errors. This can occur if the proband visits a clinic on multiple occasions and provides different family histories, or if the clinician enters multiple hypothetical versions of the pedigree to assess the risk under varying scenarios. To mimic these situations, we generate a random number of errors for each duplicate entry from a Poisson distribution with rate parameter 1.5 (average of 1.5). To produce the errors, we randomly select variables in the family history information. For binary variables, we induce the error by changing the variable to its opposite value (from 0 to 1, and from 1 to 0). For continuous variables, we randomly choose an integer from 0 to 9 to add to the existing value.

Lastly, for each of the two scenarios we generate an external data set without duplicates. This data set will be used for training and validation, as explained below.

\subsection{Model training and validation}
\label{subsec:motiv_model}

We explore the situation of training a logistic regression classifier predicting whether or not an individual carriers at least one of the two mutations. We include the following covariates in the model: indicator of proband having breast cancer, indicator of proband having ovarian cancer, proportion of relatives with breast cancer, proportion of relatives with ovarian cancer, and the proband's current age (age at time of data collection). We examine three scenarios: (1) using Monte Carlo cross-validation to randomly select half of the duplicated (augmented) data set to train the model, then evaluate model performance on the other half of the duplicated data; (2) training the model on the same training set as in scenario (1), but evaluating on a bootstrap sample of the external data set without duplicates; and (3) training on the external data set without duplicates and evaluating on a bootstrap sample of the duplicated data set. In all three scenarios, we repeat the process 100 times, with randomization coming from the random training/testing split in scenario (1) and from the bootstrap sampling in scenarios (2) and (3).

\subsection{Results}
\label{subsec:motiv_results}

We evaluate model performance using three metrics: (1) the ratio of the number of observed events to the number of expected events (O/E); (2) the area under the receiver operating characteristic curve (AUC); and (3) the root Brier score (rBS) \citep{steyerberg2010assessing}. Results for the data set with allele frequency 0.01 are displayed in Figure~\ref{fig:sim_effect_af001}. Analogous results for the data set with allele frequency 0.1 are displayed in Supplementary Figure~\ref{sfig:sim_effect_af01}.

\begin{figure}
    \centering
    \includegraphics[width=\textwidth]{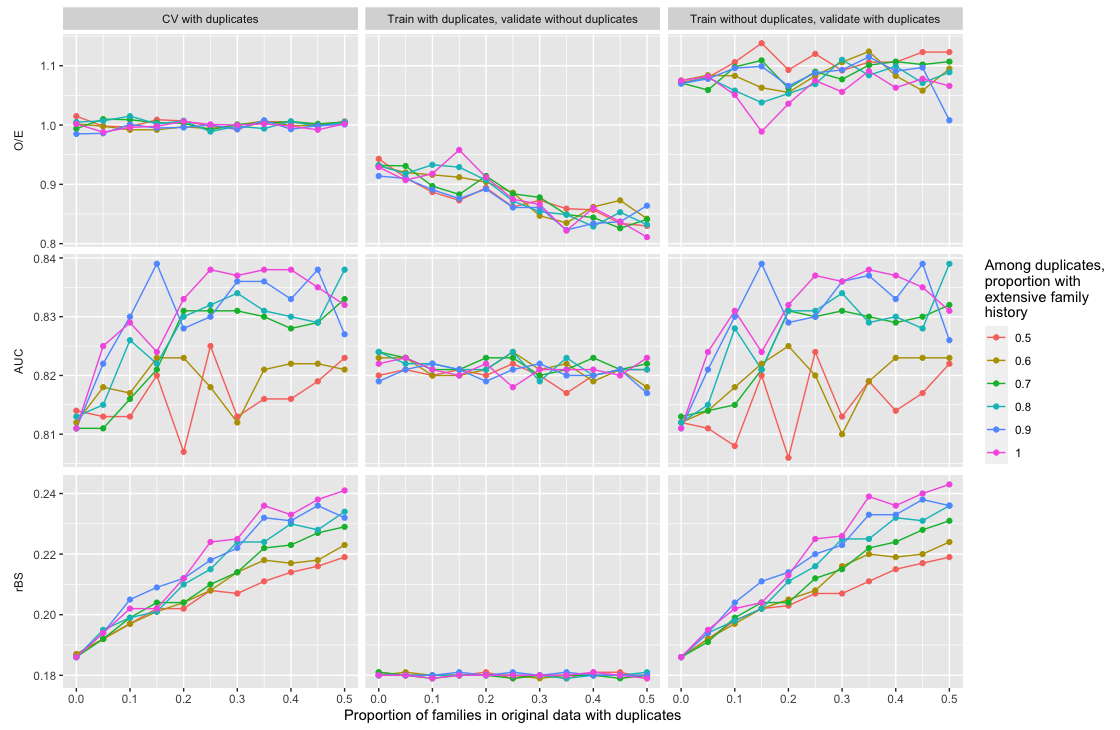}
    \caption{Performance metrics for the logistic regression model under three training/validation scenarios. The first column presents results when using Monte Carlo cross-validation on the duplicated data. The second column presents results when training on the same training set (with duplicates) as in the first column and validating on external data without duplicates. The third column presents results when training on the external data without duplicates and validating on the duplicated data. Carrier statuses are generated with \textit{BRCA1} and \textit{BRCA2} allele frequencies of 0.01. Extensive family history is defined as having at least 90th percentile of the proportion of family members with breast or ovarian cancer. Each metric is averaged over the 100 iterations of Monte Carlo cross-validation or bootstrap sampling. See Section~\ref{subsec:motiv_results} for the definitions of the three metrics.}
    \label{fig:sim_effect_af001}
\end{figure}

We see across all three scenarios that having duplicates in the training and/or validation data can impact performance metrics (Figure~\ref{fig:sim_effect_af001}). When using Monte Carlo cross-validation on the duplicated data, we see that AUC values become artificially inflated as the proportion of duplicates increase, while the rBS values increase as well. When training on the duplicated  data and evaluating on the external data without duplicates, we see that the O/E tends further from 1 as the proportion of duplicates increases. When training on the external data without duplicates and validating on the duplicated data, we see that the AUC values increase and rBS values increase as the proportion of duplicates increase. Thus, we see that duplicates in the data impact AUC and rBS values when evaluating on data with duplicates, and impact O/E when training on data with duplicates. We see similar results when using an allele frequency of 0.1 (Supplementary Figure~\ref{sfig:sim_effect_af01}). The potential impact of duplicates provides a motivation to develop methods to deduplicate pedigree data for use in both training and validation data sets.

\section{Methods} \label{sec:methods}

% \subsection{Algorithm Description}
We introduce the SNIP algorithm, an adaptation of the sorted neighborhood method in \cite{hernandez1995merge}. We define the input to the algorithm as a data matrix containing pedigrees, where each row represents an individual, such as in Table~\ref{tab:example}. For each individual (row), we require the following column variables: (1) a family ID that is common for all members of the same family, (2) an individual ID that is unique for each individual in the family, (3) a mother ID that indicates the individual ID of the mother, and (4) a father ID that indicates the individual ID of the father. The mother ID and father ID variables are coded as NA if the individual is a founder of the pedigree. In addition to these minimum requirements, the user may provide additional column variables that help distinguish between the families, such as demographic variables, information on phenotypes and genotypes, information on the clinic, date of the entry, etc. These variables will be used in the sort key as described in Section~\ref{subsec:search}.

\begin{figure}
    \centering
    \includegraphics[width=\textwidth]{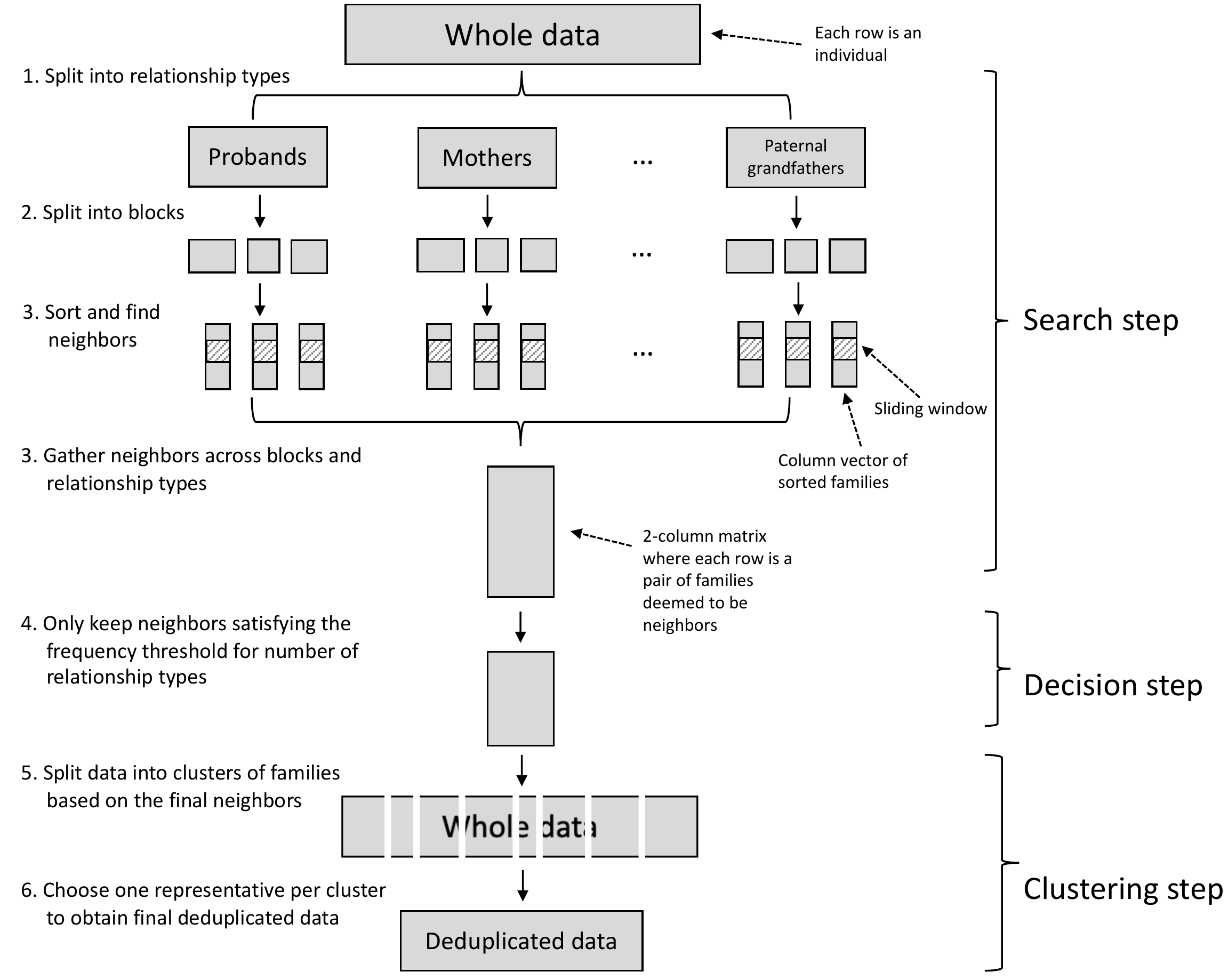}
    \caption{Flowchart of the SNIP algorithm.}
    \label{fig:flowchart}
\end{figure}

The algorithm can be formulated in three steps: (1) determining which families to compare (search step), (2) determining if the comparison in fact represents the same family (decision step), and (3) choosing a representative family from each cluster of potential duplicate families (clustering step). There are distinct issues affecting the performance of each step, motivated by our real data application. In the search step, we assume that computing all pairwise comparisons between all families is infeasible (the Risk Service data contains 317,307 families). Therefore, we employ blocking and sorted neighborhood methods to efficiently restrict the search space such that any comparisons being made are more likely to represent the same two families than a pair drawn at random. The second step, the decision step, is also challenging since we assume that there are no labels in the dataset (the Risk Service data is unlabeled), and hence are limited to unsupervised learning techniques. The third step, the clustering step, also presents the problem of determining the ``best" family (the correct one, if duplicates contain errors, or the best representative) for each cluster of families considered to be the same.
Details on the three steps are provided below. A flowchart illustrating the algorithm is provided in Figure~\ref{fig:flowchart}, and pseudocode for the algorithm is provided in Algorithm~\ref{alg:pseudo}.

\subsection{Search Step} \label{subsec:search}

In the search step, we compile a list of two-family comparisons which are more likely to be duplicates to serve as potential duplicate candidates. We first partition the input matrix into separate matrices for each prespecified core family member type, which can be determined from the mother ID and father ID. These core family member types, such as probands, mothers, paternal grandfathers, etc., are family member types that are included in all families in the data.

If the data set consists of variables of high reliability, we can further partition each family member matrix into blocks based on these variables, where families in different blocks are not considered as potential duplicates. For example, suppose duplicates are most likely to result from clinicians inputting edited versions of the same pedigree for the purposes of assessing differences in risk. In such a case, a variable such as the IP address of the computer in which the data was entered could be used as a blocking variable. This blocking process reduces the search space and thereby increases the efficiency of the search process.

The observations within the blocks are then sorted according to a randomly generated key and all two-record comparisons within a sliding window are made. The idea is that families that are near each other after the sorting should be more similar and hence stronger candidates to be duplicates. The user supplies the list of available variables for the sort key selection. If certain key variables are sex-specific (such as sex-specific cancers), the user may provide a subset of male-only or female-only variables so that the algorithm will not choose key variables of the opposite sex of the relative type (for example, the user can specify that ovarian cancer status will only be used for female relative types like mothers). The sort key, whose length is specified by the user, is generated by randomly selecting variables in the data, using the standard deviations of the variables as sampling weights. Variables with larger standard deviations are more informative in differentiating between families and hence are more highly weighted. Alternatively, the user may supply their own weights.

The sort key variables are selected one at a time, so that variables with higher weights are more likely to be selected first. We then sort each block according to the sort key variables, starting with the first variable and continuing with the subsequent ones. Each family is then compared with the subsequent families in the sorted block, where we allow the user to choose the value of a sliding window parameter which indicates the number of families to which it is compared. The process of generating sort keys and sorting can be repeated to increase the number of comparisons between families, with the number of iterations specified by the user. This search step curates a list of potential comparisons and provides a two-column matrix of two-family comparisons, with each row containing the two family IDs for the two families. Finally, we repeat the procedure for the remaining core family member types, resulting in separate matrices of candidate duplicate pairs for each core family member type.

\subsection{Decision Step} \label{subsec:decision}

After compiling the two-family comparison candidates for each core family member type and each block, we then decide if each of the two families being compared is a duplicate pair. To do so, we calculate the intersection score, defined as the number of the core family member types in which the two families are duplicate candidates (the number of times the two family IDs are found as a row in the matrices of candidate duplicate pairs). Only family pairs reaching a user-specified threshold for the intersection score (e.g., 5 out of 7 core family member types) are considered duplicate pairs.

\subsection{Clustering Step} \label{subsec:cluster}

After collecting the final matrix of duplicate pairs, we then complete the process of deduplicating the original data set. We organize the list of families in the original data into equivalence classes, or clusters, where each cluster consists of families that were considered as duplicate pairs. Hence, if family 1 is considered a duplicate of family 2 and family 2 is considered a duplicate of family 3, then all three families are in the same cluster. This partitioning is achieved in R using the \texttt{components} function in the \texttt{igraph} package \citep{csardi2006igraph}, which is typically used to obtain maximal connected components in graphs. We then choose a representative family for each equivalence class based on a user-supplied priority variable that sorts the families in each cluster. For example, if the user supplies entry dates for each family, we could choose the family with the oldest entry date. Selecting the oldest entry date would be a reasonable choice in the scenario where duplicate entries are altered pedigrees inputted by clinicians for the purpose of conducting a sensitivity analysis of the family's risk.

\begin{algorithm}
    \begin{scriptsize}
    \SetKwInOut{Input}{input}\SetKwInOut{Output}{output}
    \Input{\texttt{data}, \texttt{keyVars}, \texttt{keyWeights}, \texttt{blockVars}, \texttt{numIter}, \texttt{window}, \texttt{keyLength}, \texttt{threshold}, \texttt{priorityVar}, \texttt{priorityVarMin}}
    \Output{\texttt{dataDedup}}
    \texttt{relTypes} $\leftarrow$ (probands, mothers, fathers, maternalGrandmothers, maternalGrandfathers, paternalGrandmothers, paternalGrandfathers)\;
    \texttt{neighborsAll} $\leftarrow$ Empty matrix\;
    \tcc{Search Step:}
    \For{$i=1$ \KwTo length(\texttt{relTypes})}{
        \texttt{dataRelative} $\leftarrow$ Submatrix of \texttt{data} where the relative type is the $i$th \texttt{relType}\;
        \texttt{blocks} $\leftarrow$ Matrix where each row is one of the unique combinations of \texttt{blockVars} in \texttt{dataRelative}\;
        \texttt{neighborsRelative} $\leftarrow$ Empty matrix\;
        \For{$j=1$ \KwTo number of rows of \texttt{blocks}}{
            \texttt{dataBlock} $\leftarrow$ Subset of \texttt{dataRelative} where \texttt{blockVars} match \texttt{blocks}[j, ]\;
            \For{$k=1$ \KwTo \texttt{numIter}}{
                \texttt{key} $\leftarrow$ Random subset of \texttt{keyVars} of length \texttt{keyLength}, with sampling weights \texttt{keyWeights}\;
                \texttt{dataBlockSorted} $\leftarrow$ Column vector of the family IDs of \texttt{dataBlock}, sorted by the variables in \texttt{key} (first sorted by the first variable in \texttt{key}, then by the second, etc.)\;
                \For{$l=2$ \KwTo \texttt{window}}{
                    \texttt{dataBlockSorted} $\leftarrow$ Concatenate \texttt{dataBlockSorted} by column with the column vector of the first column of \texttt{dataBlockSorted} with the first $l-1$ entries removed, and NAs appended at the end\;
                }
                \texttt{neighbors} $\leftarrow$ First two columns of \texttt{dataBlockSorted}\;
                \For{$l=2$ \KwTo \texttt{window}}{
                    \texttt{neighbors} $\leftarrow$ Concatenate \texttt{neighbors} by row with the matrix consisting of the first and $(l+1)$th columns of \texttt{dataBlockSorted}\;
                }
                \texttt{neighbors} $\leftarrow$ Submatrix of \texttt{neighbors} only consisting of its unique rows\;
                \texttt{neighborsRelative} $\leftarrow$ Concatenate \texttt{neighborsRelative} with \texttt{neighbors} by row\;
                \texttt{neighborsRelative} $\leftarrow$ Submatrix of \texttt{neighborsRelative} consisting of its unique rows\;
            }   
        }
        \texttt{neighborsAll} $\leftarrow$ Concatenate by row \texttt{neighborsAll} with \texttt{neighborsRelative}
        % \texttt{neighborsAll}[[i]] $\leftarrow$ Concatenate \texttt{neighborsAll}[[i]] with column vector of $i$ repeated\;
    }
    \texttt{neighborsUnique} $\leftarrow$ Matrix of unique rows of \texttt{neighborsAll}, concatenated by column with a column vector indicating the number of instances each row was found in \texttt{neighborsAll}\;
    \tcc{Decision Step:}
    \texttt{neighborsThreshold} $\leftarrow$ Submatrix of \texttt{neighborsUnique} where the 3rd column is $\geq$ \texttt{threshold}\;
    \tcc{Clustering Step:}
    \texttt{clusters} $\leftarrow$ List of equivalence classes, where two families are ``related" if the pair is a row in \texttt{neighborsThreshold}. Each element of \texttt{clusters} is a vector of family IDs where each family is ``related" to each other.\;
    \eIf{priorityVarMin == \texttt{TRUE}}{
        \texttt{clusterReps} $\leftarrow$ Vector of family IDs, where the $i$th element is the family ID of the family in the $i$th element of \texttt{clusters} with the minimum value of \texttt{priorityVar}\;
    }{
        \texttt{clusterReps} $\leftarrow$ Vector of family IDs, where the $i$th element is the family ID of the family in the $i$th element of \texttt{clusters} with the maximum value of \texttt{priorityVar}\;
    }
    \texttt{dataDedup} $\leftarrow$ Submatrix of \texttt{data} where the family IDs are in \texttt{clusterReps}\;
 \end{scriptsize}
 \caption{Pseudocode for the SNIP algorithm.}
 \label{alg:pseudo}
\end{algorithm}

\section{Simulations} \label{sec:simulations}

\subsection{Design} \label{subsec:simdes}

To assess the performance of SNIP under different parameter specifications, we conduct a simulation study. We generate data in a similar way to the approaches used in Section~\ref{sec:motiv}. For these simulations, we add additional genes and phenotypes to match the ones included in the Risk Service data in Section~\ref{sec:dataapp}. We generate mutation statuses for \textit{BRCA1}, \textit{BRCA2}, \textit{MLH1}, \textit{MSH2}, \textit{MSH6}, and \textit{CDKN2A}, and generate cancer statuses and ages for breast, ovarian, colorectal, endometrial, pancreatic, and melanoma cancers. We use mutation allele frequencies and cancer penetrances from PanelPRO. We generate 5000 families using the same data generation procedure as in Section~\ref{subsec:motiv_gen}. We then randomly select 500 (10\%) families to have duplicates. Out of these 500 families, we select 100 to have 1 duplicate, 100 to have 2, 100 to have 3, 100 to have 4, and 100 to have 5. Thus, in total we have 1500 families in the data that are duplicates of an original family. We use the same process of creating duplicates as in Section~\ref{subsec:motiv_gen}. We run SNIP on both the data set with the 1500 duplicate families (total 6500 families) and the original data set without the duplicate families (total 5000 families).

Although families are generated with many member types, we consider 7 family member types as core: proband, mother, father, maternal grandmother, maternal grandfather, paternal grandmother, and paternal grandfather. These core family member types are chosen to match the ones chosen in the Risk Service data in Section~\ref{sec:dataapp}. We consider the following potential sort key variables: (binary) affection statuses for each cancer, ages of each cancer (current age if unaffected), and family size. The affection statuses of ovarian and endometrial cancers are female-specific and are thus excluded from the list of key variables for the male relative types.

We explore the following parameter configurations. For the sliding window size (number of neighbors to which we compare each family), we use 5, 10, 15, and 20. For the number of sort key iterations, we use 1, 5, and 9. For the length of each sort key, we use 3, 10, 20, and 29 (the maximum given our available sort key variables). For the intersection score threshold, we use all possibilities: 1, 2, \dots, 7. Thus, we run SNIP for 336 parameter configurations. Finally, we repeat the entire algorithm 5 times for each parameter configuration to account for variability across repetitions.

\subsection{Results} \label{subsec:simres}

We evaluate performance by examining the final deduplicated data set. The deduplicated data set can be seen as a partition of the families, where families in a cluster are considered duplicates of each other. Thus we can compare the algorithm partition (the partition produced from the algorithm) with the true partition. We use three metrics: pairwise $F_1$ \citep{manning2008introduction}, cluster $F_1$ \citep{huang2006efficient}, and generalized merge distance (GMD) \citep{menestrina2010evaluating}. Both $F_1$ measures combine notions of precision and recall by using the harmonic mean. Pairwise $F_1$ considers pairs of families within the same cluster and assesses the overlap between these pairs in the algorithm partition and the true partition. Cluster $F_1$ considers the entire clusters in the partition, and assesses the overlap between the clusters in the two partitions. Hence they diverge in their strictness on exactness of cluster overlap. Pairwise $F_1$ compares the individual families within clusters, whereas cluster $F_1$ requires the entire clusters to be exactly equal. GMD quantifies a notion of distance between two partitions, measuring the number of cluster merges and splits required to transform one partition to the other. We aim for higher values of pairwise and cluster $F_1$ and lower values of GMD. \cite{menestrina2010evaluating} provide a detailed overview of all three metrics.

\begin{figure}
    \centering
    \includegraphics[width=\textwidth]{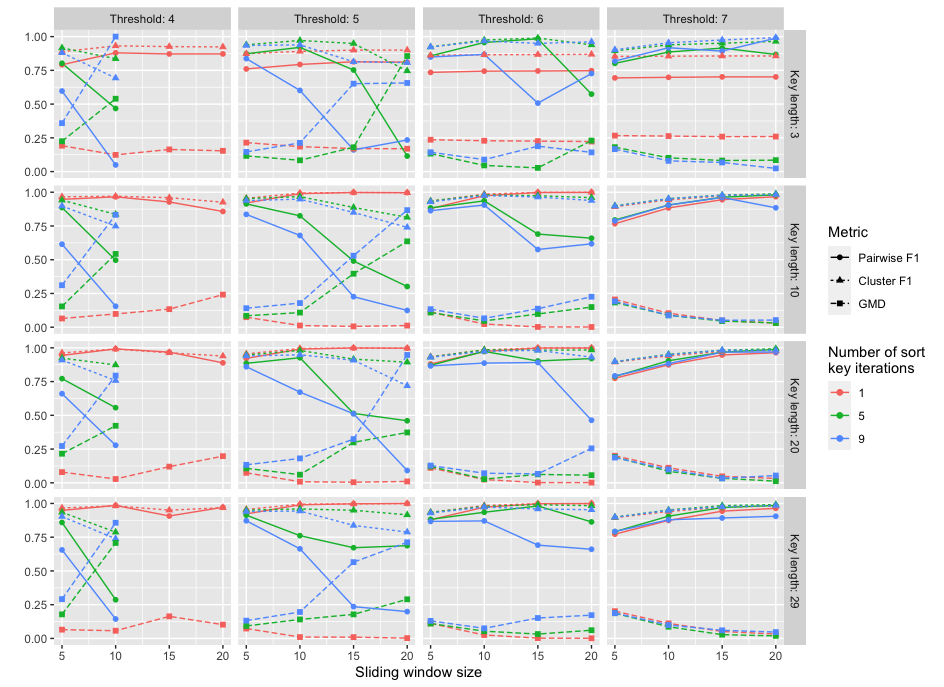}
    \caption{Comparison of the three performance metrics when applying SNIP on the deduplicated data sets in the simulation for different parameter configurations. All quantities are averaged over the 5 iterations for each parameter configuration.  Metrics were only computed when the algorithm partition had at least half of the number of clusters as the true partition, and hence some metrics are missing. Threshold is the required number of core family member types for which a pair of families needs to be considered candidate duplicates. The number of sort key iterations is the number of randomly generated sort keys used to compile the list of candidate duplicate pairs. The sliding window size is the number of adjacent families to which each family is compared after sorting using the sort key.}
    \label{fig:sim_metrics}
\end{figure}

\begin{figure}
    \centering
    \includegraphics[width=\textwidth]{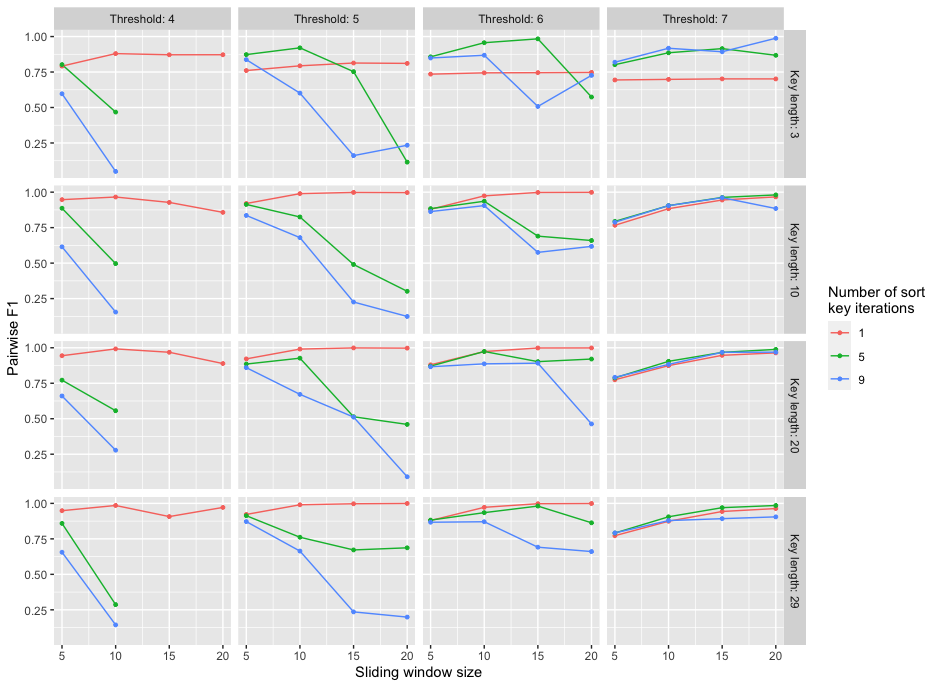}
    \caption{Average pairwise $F_1$ values (see Section~\ref{subsec:simres}) for various parameter configurations in the simulation in Section~\ref{sec:simulations}.}
    \label{sfig:sim_pf1}
\end{figure}

\begin{figure}
    \centering
    \includegraphics[width=\textwidth]{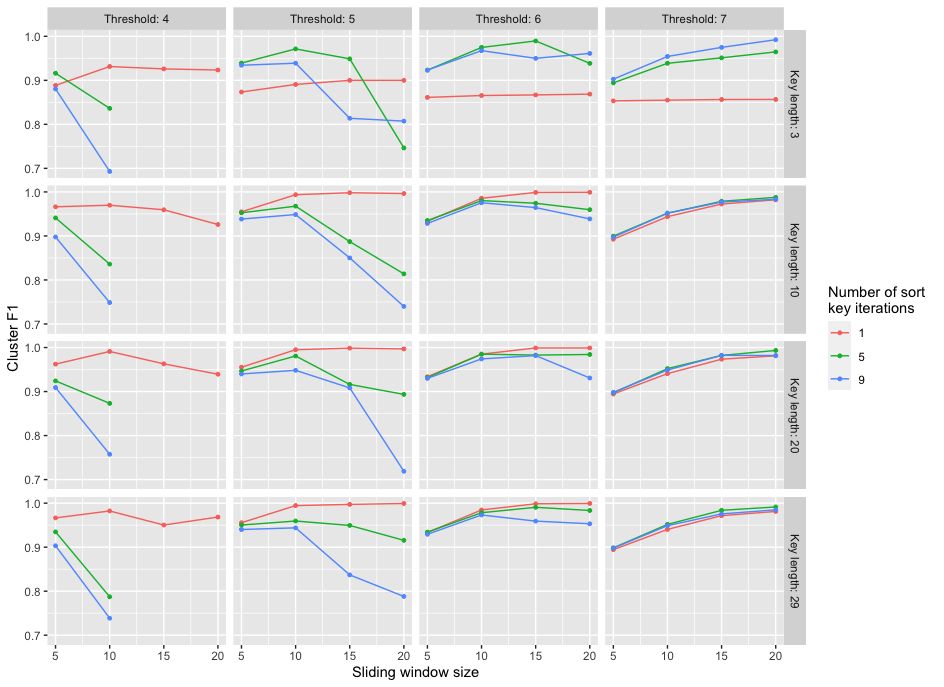}
    \caption{Average cluster $F_1$ values (see Section~\ref{subsec:simres}) for various parameter configurations in the simulation in Section~\ref{sec:simulations}.}
    \label{sfig:sim_cf1}
\end{figure}

\begin{figure}
    \centering
    \includegraphics[width=\textwidth]{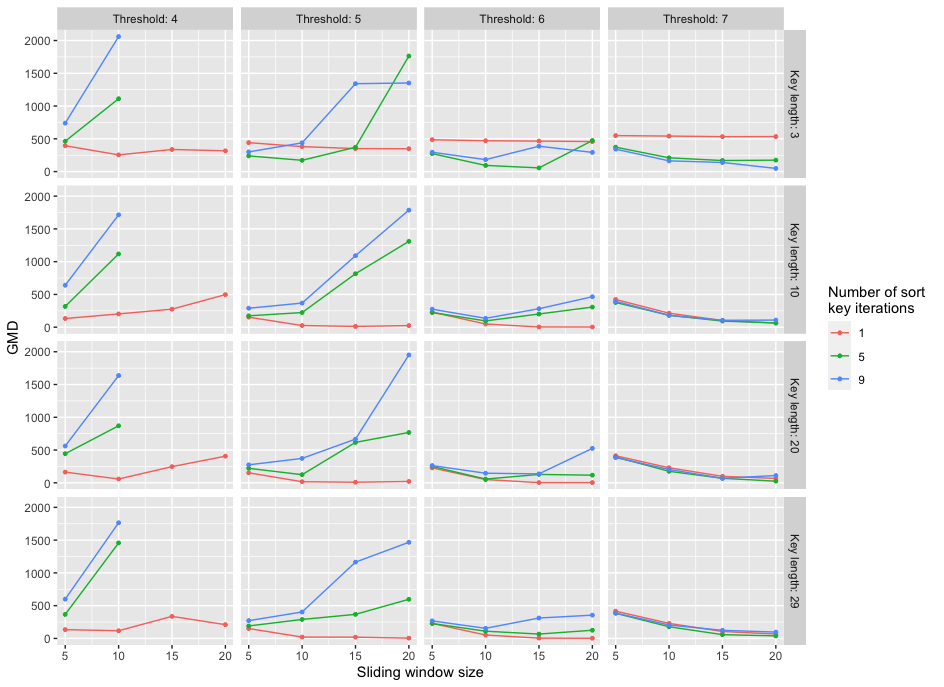}
    \caption{Average GMD values (see Section~\ref{subsec:simres}) for various parameter configurations in the simulation in Section~\ref{sec:simulations}.}
    \label{sfig:sim_gmd}
\end{figure}

Figure~\ref{fig:sim_metrics} provides results for all three metrics for different parameter configurations. For computational efficiency purposes, we only computed metrics when the parameter configuration resulted in an algorithm partition that had at least half of the number of clusters as the true partition, since algorithm partitions not satisfying this requirement perform poorly and can cause long runtimes to calculate the metrics. Hence some parameter configurations for the threshold of 4 are missing from the plot. We also omitted thresholds of 1, 2, and 3 because most of the resulting algorithm partitions had too few clusters. Any parameter configuration with missing values can be treated as one which would have performed poorly in all three metrics. In general, lower thresholds lead to more families in each cluster, and higher thresholds are more stringent and lead to fewer families in each cluster. For the simulated data set, thresholds of 5 and 6 performed best, as they provided a good balance between the two extremes. In addition, we notice that SNIP seemed to perform best when using large sliding window sizes, small number of sort key iterations, and large key lengths. Larger sliding windows increase the number of neighbors considered, and larger key lengths use more variables to search for the neighbors. However, increasing the number of sort key iterations excessively increases the number of neighbors, worsening the performance for smaller thresholds. Overall, there seemed to be a mostly monotonic relationship between these parameters and the three metrics. All three metrics were highly correlated and provided similar results. Supplementary Table~\ref{stab:sim_params} also provides the 10 parameter configurations that provided the lowest GMD values, and Supplementary Figures~\ref{sfig:sim_pf1}, \ref{sfig:sim_cf1}, and \ref{sfig:sim_gmd} provide plots of the three metrics individually.

\begin{figure}
    \centering
    \includegraphics[width=\textwidth]{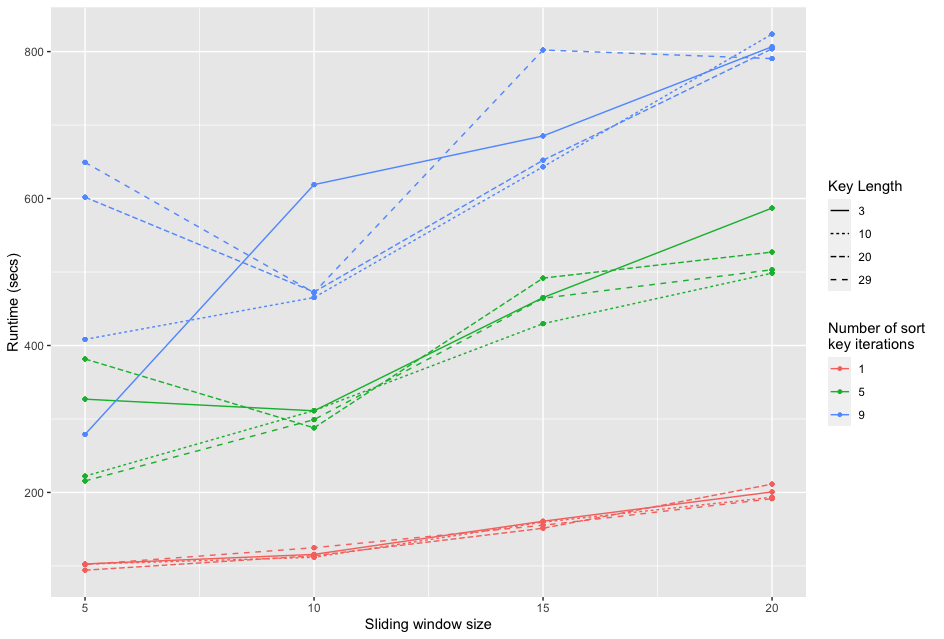}
    \caption{Average runtimes for each parameter configuration in the simulation in Section~\ref{sec:simulations}.}
    \label{sfig:sim_runtime}
\end{figure}

Another measure of performance is the algorithm efficiency as measured by the runtime. Supplementary Figure~\ref{sfig:sim_runtime} plots the runtimes for each parameter configuration. For a given sliding window size, key length, and number of sort key iterations, the neighbors are determined for each threshold simultaneously, so runtime is not dependent on threshold. In general, runtimes may vary substantially depending on the computer used to run the algorithm; for the work in this paper, we used the FASRC Cannon cluster supported by the FAS Division of Science Research Computing Group at Harvard University. For our implementation, we see that runtime mostly increases with the number of sort key iterations and sliding window size, as increasing these parameters increases the number of neighbors considered. Conversely, the key length does not impact the number of neighbors considered, and hence runtimes are largely unaffected by the key length.

\begin{figure}
    \centering
    \includegraphics[width=\textwidth]{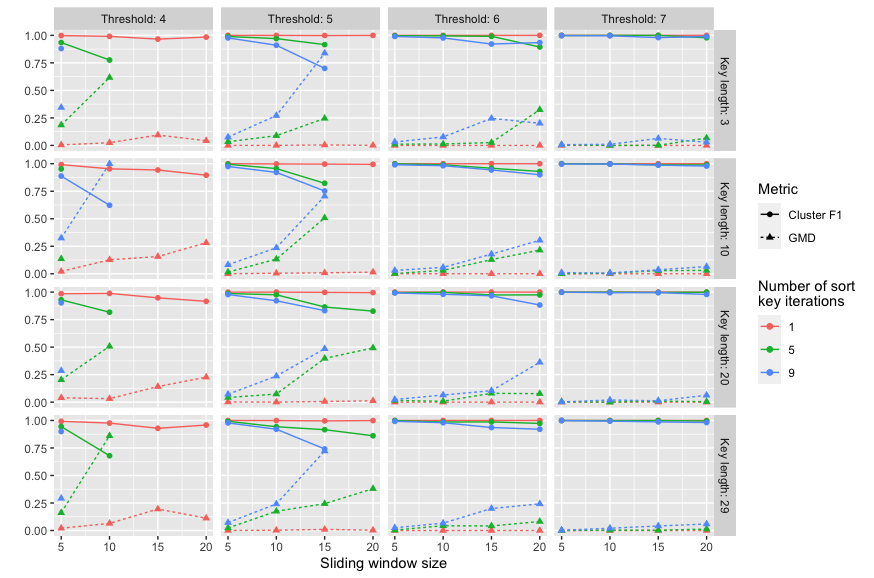}
    \caption{Performance metrics for the deduplicated data sets in the simulated data sets without true duplicates. All quantities are averaged over the 5 iterations for each parameter configuration.  Metrics were only computed when the algorithm partition had at least half of the number of clusters as the true partition, and hence some metrics are missing.}
    \label{sfig:sim_metrics_nodup}
\end{figure}

SNIP performed well even when the data set did not contain any true duplicates. Supplementary Figure~\ref{sfig:sim_metrics_nodup} provides cluster $F_1$ and GMD values for various parameter configurations. Pairwise $F_1$ values are all indeterminate in this case, since the true partition does not contain pairs. When we used one sort key iteration and a threshold of 7, the deduplicated data sets were all identical to the original data set. When we used one sort key iteration and a threshold of 6, the deduplicated data sets at most removed 4 families (out of 5000). This provides evidence that, with the right parameters, SNIP may potentially be used on any data set, even without prior belief that the data set contains duplicates.

\section{Data Application} \label{sec:dataapp}

\subsection{Risk Service Data} \label{subsec:riskservice}

We apply SNIP to data collected through the Risk Service, a web interface developed by the HughesRiskApps group (Massachusetts General Hospital) and the BayesMendel lab (Dana-Farber Cancer Institute) to provide clinicians with family-based risk calculations for multiple cancer risk prediction models \citep{risk_service}. The software is used at many clinics, and automatically stores pedigree data when clinicians collect family history information from probands during clinical visits. The information is then used to predict risk of cancer susceptibility. Since the data entries are used as inputs in risk prediction models via risk assessment software with pre-specified fields, the data is relatively clean with standardized values for the different variables.

The data set is being continually updated, as clinicians use these tools on a regular basis. The current extraction of the data is from January 2018 and contains 317,307 families with probands who consented to the use of their data for research. Probands are asked to provide family history for both first-degree and second-degree family members. Each proband is required to provide information on at least the 7 core family member types used in Section~\ref{sec:simulations} for the simulation; however, much of the information on relatives is missing, such as the ages. As in the simulations in Section~\ref{sec:simulations}, the family history includes current ages; statuses and ages of diagnosis for breast, ovarian, colorectal, endometrial, pancreatic, and melanoma cancers; genetic testing results if available for \textit{BRCA1}, \textit{BRCA2}, \textit{MLH1}, \textit{MSH2}, \textit{MSH6}, and \textit{CDKN2A}. The data also contain additional variables, such as intervention status and age for prophylactic mastectomies and oophorectomies; Ashkenazi Jewish status; tumor marker status for ER, PR, and HER2; twin status; presence and location of microsatellite instability for colorectal tumors; date and time of the entry; and IP address of the computer used to input the information.
Most of the probands in the data are being evaluated for breast cancer risk, resulting in an enriched number of breast and ovarian cancer cases among family members as well as a high proportion of female probands (99.12\%).
Table~\ref{tab:risk_service} (first column) provides summary characteristics of the families.

The data may contain duplicate families due to multiple visits by the same proband or multiple entries in the same visit. Clinicians may enter different family histories during the same visit for the purpose of risk comprehension--for example, they may alter the family history to provide risk in an alternative hypothetical setting. In addition, the data likely contain automatically generated test families, as there are large groups of similar families entered at the same IP address in a short time window.  These traits result in a data set with potentially numerous duplicates and hence is ideal for applying SNIP. However, the data are de-identified and the true duplicate statuses are unknown; thus, we cannot use the data for validating our algorithm but only for exploring and understanding the behavior of the algorithm in an application on real data.

\begin{table}[]
    \centering
    \begin{tabular}{rrr}
         \toprule
         & Whole data & Deduplicated data$^*$ \\
         \midrule
         \multicolumn{3}{l}{\textbf{General}} \\
         Number of families & 317,307 & 212,787 \\
         Average family size & 13.02 & 13.24 \\
         Proband current age & 52.51 & 53.98 \\
         \\
         \multicolumn{3}{l}{\textbf{Proband cancer history}} \\
         \multicolumn{3}{l}{Proportion with:} \\
         Breast & 8.57\% & 9.87\% \\
         Ovarian & 0.71\% & 0.82\% \\
         Colon & 0.65\% & 0.73\% \\
         Endometrial & 0.72\% & 0.83\% \\
         Pancreatic & 0.07\% & 0.08\% \\
         Melanoma & 1.26\% & 1.56\% \\
         \\
         \multicolumn{3}{l}{\textbf{Relatives cancer history}} \\
         \multicolumn{3}{l}{Proportion with:} \\
         Breast & 6.62\% & 6.92\% \\
         Ovarian & 0.83\% & 0.93\% \\
         Colon & 1.84\% & 2.20\% \\
         Endometrial & 0.41\% & 0.47\% \\
         Pancreatic & 0.52\% & 0.61\% \\
         Melanoma & 0.64\% & 0.75\% \\
         \\
         \multicolumn{3}{l}{\textbf{Proband genetic testing}} \\
         \multicolumn{3}{l}{Number tested (number with pathogenic mutation)} \\
         \textit{BRCA1} & 9422 (363) & 4823 (184) \\
         \textit{BRCA2} & 9451 (385) & 4835 (188) \\
         \textit{MLH1} & 4414 (22) & 1552 (15) \\
         \textit{MSH2} & 4408 (31) & 1549 (11) \\
         \textit{MSH6} & 4396 (26) & 1526 (7) \\
         \textit{CDKN2A} & 4 (1) & 2 (0) \\
         \\
         \multicolumn{3}{l}{\textbf{Relatives genetic testing}} \\
         \multicolumn{3}{l}{Number tested (number with pathogenic mutation)} \\
         \textit{BRCA1} & 6453 (1119) & 3549 (541) \\
         \textit{BRCA2} & 6709 (1240) & 3752 (666) \\
         \textit{MLH1} & 1282 (58) & 666 (37) \\
         \textit{MSH2} & 1324 (99) & 670 (48) \\
         \textit{MSH6} & 1321 (125) & 670 (65) \\
         \textit{CDKN2A} & 7 (4) & 3 (2) \\
         \bottomrule
    \end{tabular} \\
    $^*$ Sliding window size of 20, threshold of 7
    \caption{Descriptive statistics for the Risk Service data. The first column is for the entire data, and the second column is for the deduplicated data after running SNIP with a sliding window size of 20 and a threshold of 7.}
    \label{tab:risk_service}
\end{table}

We apply SNIP on the data set, using parameter values based on the simulation results in Section~\ref{subsec:simres}. We use one sort key iteration as the simulation results provided evidence that one iteration is sufficient. We use a key length of 31, the maximum value, since the simulation results provided evidence that large key lengths were optimal. 31 is the maximum value since there are only 31 possible key variables: affection statuses and ages for breast, ovarian, colorectal, endometrial, pancreatic, and melanoma skin cancers (current ages are used if the individual does not have the cancer); intervention statuses and ages for prophylactic mastectomies and oophorectomies; Ashkenazi Jewish status; genetic test results for \textit{BRCA1}, \textit{BRCA2}, \textit{MLH1}, \textit{MSH2}, \textit{MSH6}, and \textit{CDKN2A}; tumor marker statuses for ER, PR, and HER2; twin status, presence and location of microsatellite instability for colorectal tumors; and family size. Breast cancer status has three possible outcomes: no cancer, unilateral breast cancer, and bilateral breast cancer. We have two variables for breast cancer age: age of first breast cancer and age of contralateral breast cancer. We explore results for all 7 thresholds. We use the IP address of the clinician as a blocking variable, since clinicians recording multiple entries at the same visit will have the same IP address, and families are assumed to visit the same clinician across different visits. Finally, we use three sliding window sizes--5, 10, and 20--to assess the differences in results when considering more or fewer neighbors. As in the simulations, we used the Harvard FASRC Cannon cluster to run SNIP.

\subsection{Results} \label{subsec:datares}

\begin{figure}
    \centering
    \includegraphics[width=\textwidth]{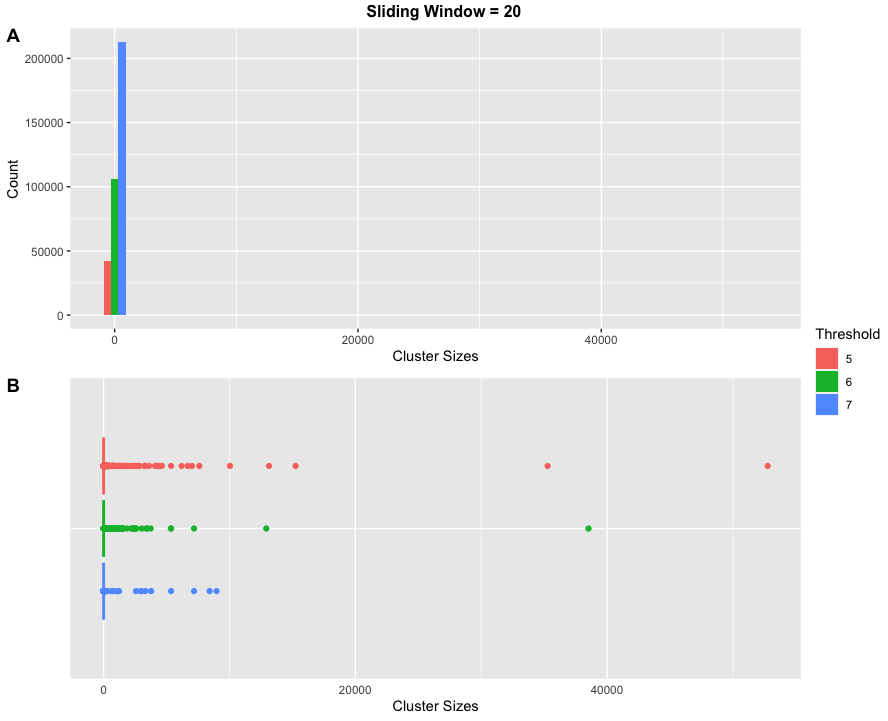}
    \caption{Cluster sizes and counts for the Risk Service Data, with a sliding window size of 20 and thresholds of 5, 6, and 7. Subfigure (A) provides a histogram, with cluster sizes $>1$ invisible because of the skewness of the distribution. Subfigure (B) provides a boxplot that shows the large cluster sizes.}
    \label{sfig:cluster_20}
\end{figure}

The Risk Service data contain many more families compared to the simulations, and the choices of parameters may impact the results in different ways. The runtimes for the sliding window sizes of 5, 10, and 20 were 1107, 1937, and 4649 seconds, respectively, while the memory usages were 8.77, 12.83, and 22.51 gigabytes, respectively. A frequency table for the cluster sizes when using a threshold of 7, for each of the three sliding window sizes, is provided in Table~\ref{tab:cluster_7}. Plots for the three sliding window sizes, using thresholds of 5, 6, and 7, are provided in Supplementary Figures~\ref{sfig:cluster_5}, \ref{sfig:cluster_10}, and \ref{sfig:cluster_20}, respectively.

\begin{table}[ht]
\centering
\begin{tabular}{rrrrrrrr}
  \toprule
  \textbf{Cluster Size} & \multicolumn{3}{c}{\textbf{Number of clusters}} & \textbf{Cluster Size} & \multicolumn{3}{c}{\textbf{Number of clusters}} \\
  \cmidrule(lr){2-4} \cmidrule(lr){6-8}
 & \multicolumn{3}{c}{Sliding Window Size} & & \multicolumn{3}{c}{Sliding Window Size} \\ 
& 5 & 10 & 20 & & 5 & 10 & 20 \\
  \midrule
  1 & 200232 & 191517 & 177451 &  44 &   0 &   0 &   1 \\ 
    2 & 22768 & 24624 & 26256 &  45 &   0 &   0 &   2 \\ 
    3 & 3731 & 4467 & 5770 &  46 &   0 &   0 &   1 \\ 
    4 & 1085 & 1346 & 1885 &  48 &   1 &   1 &   2 \\ 
    5 & 388 & 440 & 659 &  52 &   0 &   0 &   1 \\ 
    6 & 153 & 191 & 297 &  53 &   0 &   0 &   2 \\ 
    7 &  63 &  90 & 128 &  54 &   1 &   1 &   1 \\ 
    8 &  41 &  51 &  72 &  56 &   1 &   0 &   0 \\ 
    9 &  28 &  29 &  51 &  57 &   0 &   0 &   3 \\ 
   10 &  14 &  18 &  40 &  59 &   0 &   0 &   1 \\ 
   11 &   8 &  13 &  16 &  60 &   0 &   0 &   1 \\ 
   12 &  11 &  14 &  19 &  62 &   0 &   0 &   1 \\ 
   13 &   6 &   7 &   8 &  68 &   0 &   0 &   1 \\ 
   14 &   4 &   7 &  13 &  72 &   1 &   1 &   1 \\ 
   15 &   3 &   3 &   5 &  75 &   0 &   0 &   1 \\ 
   16 &   2 &  11 &   7 &  76 &   0 &   0 &   1 \\ 
   17 &   1 &   5 &   6 &  80 &   0 &   0 &   1 \\ 
   18 &   0 &   5 &   6 &  89 &   1 &   0 &   0 \\ 
   19 &   5 &   2 &   6 & 100 &   1 &   1 &   1 \\ 
   20 &   1 &   6 &   5 & 143 &   1 &   1 &   1 \\ 
   21 &   0 &   1 &   2 & 145 &   1 &   0 &   0 \\ 
   22 &   0 &   3 &   2 & 146 &   0 &   1 &   1 \\ 
   23 &   2 &   4 &   3 & 157 &   0 &   1 &   1 \\ 
   24 &   1 &   3 &   5 & 199 &   0 &   1 &   0 \\ 
   25 &   3 &   2 &   4 & 277 &   0 &   0 &   1 \\ 
   26 &   0 &   0 &   2 & 596 &   1 &   1 &   1 \\ 
   27 &   0 &   2 &   4 & 769 &   1 &   1 &   1 \\ 
   28 &   0 &   0 &   5 & 1097 &   2 &   2 &   2 \\ 
   29 &   1 &   0 &   2 & 1228 &   1 &   1 &   1 \\ 
   30 &   1 &   1 &   2 & 2572 &   1 &   1 &   0 \\ 
   31 &   1 &   2 &   3 & 2574 &   0 &   0 &   1 \\ 
   32 &   0 &   1 &   0 & 2931 &   1 &   1 &   1 \\ 
   34 &   0 &   0 &   1 & 3021 &   1 &   1 &   1 \\ 
   35 &   0 &   0 &   1 & 3320 &   1 &   1 &   1 \\ 
   36 &   0 &   1 &   4 & 3778 &   1 &   1 &   1 \\ 
   37 &   1 &   1 &   1 & 5363 &   1 &   1 &   1 \\ 
   38 &   1 &   0 &   1 & 7190 &   1 &   1 &   1 \\ 
   41 &   0 &   0 &   1 & 8424 &   1 &   1 &   1 \\ 
   42 &   0 &   0 &   2 & 8978 &   1 &   1 &   1 \\ 
   43 &   1 &   1 &   2 &  &  &  &  \\ 
      \bottomrule
\end{tabular}
    \caption{Cluster sizes and counts for the Risk Service data identified with SNIP, with a threshold of 7 and sliding window sizes of 5, 10, and 20.}
    \label{tab:cluster_7}
\end{table}

We notice that all three sliding window sizes result in large clusters. Even when using the strictest threshold of 7, all three sliding window sizes produce clusters with as many as 8978 families. Delving deeper into these large clusters shows that these families may be automatically generated test families. Supplementary Tables~\ref{stab:summ_within_cluster_20}, \ref{stab:summ_within_cluster_10}, and \ref{stab:summ_within_cluster_5} summarize the differences between the families within clusters for the large clusters (size $\geq 100$) for a threshold of 7 and sliding window sizes of 20, 10, and 5, respectively. For all 3 sliding window sizes, all of the families within any of these large clusters have the same number of family members and consist of a proband with the same cancer status (cancer or no cancer) and the same current age, with each of the other 6 core relative types all having the same cancer status and current age. Almost all of the clusters have families with probands who are cancer-free at the same current age and relatives who are cancer-free at an unknown current age. Moreover, these families are all entered into the database within a short time frame (usually within a day), and are entered from the same IP address (by design, since we use IP address as a blocking variable). These characteristics, along with the abnormally large cluster sizes, lead us to believe that the families may be automatically generated test families. SNIP is hence seemingly able to detect these extreme clusters, even for all 3 sliding window sizes. Using smaller thresholds will group some of these clusters of test families, so a threshold of 7 most likely is optimal for this data.

Table~\ref{tab:risk_service} also provides a comparison between the full and deduplicated data sets when using a sliding window size of 20 and a threshold of 7. Overall, SNIP removed 32\% of the original families, identifying them as duplicates. We note that the families in the deduplicated data have a richer cancer history, both among probands and among relatives, perhaps due to the removal of many of the cancer-free families in the large clusters. The removal of the families in the large clusters also causes the average family size to increase, since the families in the large clusters mostly consisted of minimum-sized families of only the 7 core relative types. The differences between the original and deduplicated data sets could lead to differences in building and validating prediction models, as shown in Section~\ref{sec:motiv}.

\section{Discussion} \label{sec:discussion}

This paper introduces and implements SNIP, a fast and unsupervised learning algorithm to find and classify pairwise record comparisons as duplicates in pedigree data.
We first demonstrated the importance of deduplication through a motivating simulation. We then used an additional simulation to assess the performance of SNIP under various parameter configurations. We found that under certain paramter configurations, SNIP performed well when measured by cluster $F_1$, pairwise $F_1$, and the generalized merge distance. We also applied SNIP to the Risk Service data set and showed an ability to detect potential large clusters of automatically generated test families.

SNIP relies on several assumptions that may limit its utility in certain situations. First, it relies on the mother IDs and father IDs being correctly reported, since it partitions families into the 7 core relative types. Data entry errors may be common in large pedigree data sets, and SNIP will likely not be able to detect duplicates of families with these errors. It also assumes that blocking variables are reliable, as it does not consider pairs of families in different blocks as potential duplicates. In general, this assumption can be relaxed by using the variable as a key variable instead of a blocking variable, and using a large key weight to ensure that the variable is selected towards the beginning of the sort key. Lastly, SNIP assumes that all families in the data have the same core family members. This was the case for the Risk Service data, but is often not the case in practice. In these situations, we could require fewer family member types or use lower thresholds in the decision step.

SNIP reduces the pedigrees to the 7 core relative types, and only incorporates the rest of the pedigree through the key variables. The number of family members was used as a key variable for the simulations and the Risk Service data application. This simplification worked well in the simulations, but may not be as effective in other data-generating scenarios. The pedigrees in the simulations only included first- and second-degree relatives; however, large families with higher-degree relatives are not uncommon in pedigree data, especially in data consisting of families at high risk of hereditary diseases. Using an alternative to the intersection score which incorporates these other relatives and utilizes the pedigree structure could be more effective. We introduced an alternative which we call the greedy match score which uses all the family members; however, this scoring system was found to be ineffective in the simulations. Details about the greedy match score and its application in the simulations are provided in Supplementary Section~\ref{ssec:greedy}. Further research is necessary to design better scoring systems, such as ones similar to the strategies introduced in \cite{belin1995method} and \cite{larsen2001iterative}, and to explore specific situations where such scores would improve upon the intersection score.

Overall, the simulations provided an illustrative view of SNIP's performance and are a good introductory step. However, a more extensive simulation study could be conducted to explore additional scenarios, such as varying the following quantities: sample sizes, family sizes, pedigree structures, levels of missing data in the key variables, number of key variables, cluster sizes, errors in a duplicate, and types of errors (such as removing family members).

When applying SNIP to the Risk Service data, we chose a parameter configuration based mostly on the simulation results, despite the differences between the simulated data and the Risk Service data.
As stated above, a simulation study that more closely resembles the Risk Service data may better inform the parameter choice. In practice, it may be worthwhile to explore various parameter configurations on a given data set and examine the resulting clusters.
For the Risk Service data, we found that our choices of parameters were sufficient to uncover the seemingly automatically generated test families while maintaining reasonable runtimes. SNIP may also be used as a data exploration tool, rather than a data cleaning tool, to help the user understand the extent of duplication in the data and to provide an initial search to identify potential duplicate clusters.

We used a static sorted neighborhood method, but adaptive methods have been developed \citep{adaptive_blocking,adaptive_sn,adaptive_windows} and may be used to allow for dynamic window sizes based on the data. In this paper, we chose to start with the static sorted neighborhood method due to simplicity and the relative homogeneity of observations in typical pedigree data sets, which may lessen the advantages of adaptive approaches. Exploring adaptive approaches would be an informative potential future direction.

We used an unsupervised approach, but a supervised approach may be more effective if we have access to training data that include labels of known duplicate clusters. Numerous modeling strategies could be implemented to learn the patterns of typical duplicates and the types of common duplication errors. One potential hybrid approach could be to inject known duplicates into unlabeled data by duplicating some of the families, and then training a prediction model using the combined true and synthetic data.

Overall, the proposed SNIP algorithm in this paper provides a useful tool to detect duplicate entries in pedigree data. Duplicate families can be a subtle but serious issue when analyzing pedigree data, and identifying the duplicates is important for using the data to make valid conclusions. We hope that SNIP can be a contribution toward exploring, cleaning, and analyzing pedigree data.

\section{Acknowledgments}
We wish to thank Giovanni Parmigiani and members of the BayesMendel lab for their valuable feedback during the preparation of this manuscript. T.H. was supported by NIH grant T32 CA 009001. M.P. was supported by NIH grant CA09337. D.B. was supported by the Friends of Dana-Farber.

\clearpage

\bibliography{bib}

\begin{thebibliography}{}

\bibitem[\protect\citeauthoryear{Belin and Rubin}{Belin and
  Rubin}{1995}]{belin1995method}
Belin, T.~R. and Rubin, D.~B. (1995).
\newblock A method for calibrating false-match rates in record linkage.
\newblock {\em Journal of the American Statistical Association} {\bf 90,}
  694--707.

\bibitem[\protect\citeauthoryear{Bilenko, Kamath, and Mooney}{Bilenko
  et~al.}{2006}]{adaptive_blocking}
Bilenko, M., Kamath, B., and Mooney, R.~J. (2006).
\newblock Adaptive blocking: Learning to scale up record linkage.
\newblock In {\em Data Mining, 2006. ICDM'06. Sixth International Conference
  on}, pages 87--96. IEEE.

\bibitem[\protect\citeauthoryear{Calado, Herschel, and Leit{\~a}o}{Calado
  et~al.}{2010}]{calado2010overview}
Calado, P., Herschel, M., and Leit{\~a}o, L. (2010).
\newblock An overview of xml duplicate detection algorithms.
\newblock {\em Soft Computing in XML Data Management} pages 193--224.

\bibitem[\protect\citeauthoryear{Chen, Wang, Broman, Katki, and
  Parmigiani}{Chen et~al.}{2004}]{bayesmendel}
Chen, S., Wang, W., Broman, K.~W., Katki, H.~A., and Parmigiani, G. (2004).
\newblock Bayesmendel: an r environment for mendelian risk prediction.
\newblock {\em Statistical applications in genetics and molecular biology} {\bf
  3,} 1--19.

\bibitem[\protect\citeauthoryear{Chipman, Drohan, Blackford, Parmigiani,
  Hughes, and Bosinoff}{Chipman et~al.}{2013}]{risk_service}
Chipman, J., Drohan, B., Blackford, A., Parmigiani, G., Hughes, K., and
  Bosinoff, P. (2013).
\newblock Providing access to risk prediction tools via the hl7 xml-formatted
  risk web service.
\newblock {\em Breast cancer research and treatment} {\bf 140,} 187--193.

\bibitem[\protect\citeauthoryear{Csardi, Nepusz, et~al\mbox{.}}{Csardi
  et~al.}{2006}]{csardi2006igraph}
Csardi, G., Nepusz, T., et~al. (2006).
\newblock The igraph software package for complex network research.
\newblock {\em InterJournal, complex systems} {\bf 1695,} 1--9.

\bibitem[\protect\citeauthoryear{Dhivyabharathi and Kumaresan}{Dhivyabharathi
  and Kumaresan}{2016}]{survey_2016}
Dhivyabharathi, G. and Kumaresan, S. (2016).
\newblock A survey on duplicate record detection in real world data.
\newblock In {\em Advanced Computing and Communication Systems (ICACCS), 2016
  3rd International Conference on}, volume~1, pages 1--5. IEEE.

\bibitem[\protect\citeauthoryear{Draisbach, Naumann, Szott, and
  Wonneberg}{Draisbach et~al.}{2012}]{adaptive_windows}
Draisbach, U., Naumann, F., Szott, S., and Wonneberg, O. (2012).
\newblock Adaptive windows for duplicate detection.
\newblock In {\em Data Engineering (ICDE), 2012 IEEE 28th International
  Conference on}, pages 1073--1083. IEEE.

\bibitem[\protect\citeauthoryear{Elmagarmid, Ipeirotis, and
  Verykios}{Elmagarmid et~al.}{2007}]{survey_2007}
Elmagarmid, A.~K., Ipeirotis, P.~G., and Verykios, V.~S. (2007).
\newblock Duplicate record detection: A survey.
\newblock {\em IEEE Transactions on knowledge and data engineering} {\bf 19,}
  1--16.

\bibitem[\protect\citeauthoryear{Freedman, Slattery, Ballard-Barbash, Willis,
  Cann, Pee, Gail, and Pfeiffer}{Freedman
  et~al.}{2009}]{freedman2009colorectal}
Freedman, A.~N., Slattery, M.~L., Ballard-Barbash, R., Willis, G., Cann, B.~J.,
  Pee, D., Gail, M.~H., and Pfeiffer, R.~M. (2009).
\newblock Colorectal cancer risk prediction tool for white men and women
  without known susceptibility.
\newblock {\em Journal of clinical oncology} {\bf 27,} 686.

\bibitem[\protect\citeauthoryear{Hern{\'a}ndez and Stolfo}{Hern{\'a}ndez and
  Stolfo}{1995}]{hernandez1995merge}
Hern{\'a}ndez, M.~A. and Stolfo, S.~J. (1995).
\newblock The merge/purge problem for large databases.
\newblock {\em ACM Sigmod Record} {\bf 24,} 127--138.

\bibitem[\protect\citeauthoryear{Huang, Ertekin, and Giles}{Huang
  et~al.}{2006}]{huang2006efficient}
Huang, J., Ertekin, S., and Giles, C.~L. (2006).
\newblock Efficient name disambiguation for large-scale databases.
\newblock In {\em European conference on principles of data mining and
  knowledge discovery}, pages 536--544. Springer.

\bibitem[\protect\citeauthoryear{Ivie, Pixton, and Giraud-Carrier}{Ivie
  et~al.}{2007}]{ivie2007metric}
Ivie, S., Pixton, B., and Giraud-Carrier, C. (2007).
\newblock Metric-based data mining model for genealogical record linkage.
\newblock In {\em 2007 IEEE International Conference on Information Reuse and
  Integration}, pages 538--543. IEEE.

\bibitem[\protect\citeauthoryear{Kolb, Thor, and Rahm}{Kolb
  et~al.}{2012}]{mapreduce}
Kolb, L., Thor, A., and Rahm, E. (2012).
\newblock Multi-pass sorted neighborhood blocking with mapreduce.
\newblock {\em Computer Science-Research and Development} {\bf 27,} 45--63.

\bibitem[\protect\citeauthoryear{K{\"o}pcke and Rahm}{K{\"o}pcke and
  Rahm}{2010}]{comparison_2010}
K{\"o}pcke, H. and Rahm, E. (2010).
\newblock Frameworks for entity matching: A comparison.
\newblock {\em Data \& Knowledge Engineering} {\bf 69,} 197--210.

\bibitem[\protect\citeauthoryear{Larsen and Rubin}{Larsen and
  Rubin}{2001}]{larsen2001iterative}
Larsen, M.~D. and Rubin, D.~B. (2001).
\newblock Iterative automated record linkage using mixture models.
\newblock {\em Journal of the American Statistical Association} {\bf 96,}
  32--41.

\bibitem[\protect\citeauthoryear{Manning, Raghavan, and Sch{\"u}tze}{Manning
  et~al.}{2008}]{manning2008introduction}
Manning, C.~D., Raghavan, P., and Sch{\"u}tze, H. (2008).
\newblock {\em Introduction to Information retrieval}.
\newblock Cambridge University Press.

\bibitem[\protect\citeauthoryear{Menestrina, Whang, and
  Garcia-Molina}{Menestrina et~al.}{2010}]{menestrina2010evaluating}
Menestrina, D., Whang, S.~E., and Garcia-Molina, H. (2010).
\newblock Evaluating entity resolution results.
\newblock {\em Proceedings of the VLDB Endowment} {\bf 3,} 208--219.

\bibitem[\protect\citeauthoryear{Papadakis, Koutrika, Palpanas, and
  Nejdl}{Papadakis et~al.}{2014}]{metablocking}
Papadakis, G., Koutrika, G., Palpanas, T., and Nejdl, W. (2014).
\newblock Meta-blocking: Taking entity resolutionto the next level.
\newblock {\em IEEE Transactions on Knowledge and Data Engineering} {\bf 26,}
  1946--1960.

\bibitem[\protect\citeauthoryear{Pixton and Giraud-Carrier}{Pixton and
  Giraud-Carrier}{2005}]{pixton2005mal4}
Pixton, B. and Giraud-Carrier, C. (2005).
\newblock Mal4: 6-using data mining for record linkage.
\newblock In {\em Proceedings of the 5th annual Workshop on technology for
  family history and genealogical research}. Citeseer.

\bibitem[\protect\citeauthoryear{Pixton and Giraud-Carrier}{Pixton and
  Giraud-Carrier}{2006}]{pixton2006using}
Pixton, B. and Giraud-Carrier, C. (2006).
\newblock Using structured neural networks for record linkage.
\newblock In {\em Proceedings of the sixth annual workshop on technology for
  family history and genealogical research}.

\bibitem[\protect\citeauthoryear{Steyerberg, Vickers, Cook, Gerds, Gonen,
  Obuchowski, Pencina, and Kattan}{Steyerberg
  et~al.}{2010}]{steyerberg2010assessing}
Steyerberg, E.~W., Vickers, A.~J., Cook, N.~R., Gerds, T., Gonen, M.,
  Obuchowski, N., Pencina, M.~J., and Kattan, M.~W. (2010).
\newblock Assessing the performance of prediction models: a framework for some
  traditional and novel measures.
\newblock {\em Epidemiology (Cambridge, Mass.)} {\bf 21,} 128.

\bibitem[\protect\citeauthoryear{Tyrer, Duffy, and Cuzick}{Tyrer
  et~al.}{2004}]{tyrer2004breast}
Tyrer, J., Duffy, S.~W., and Cuzick, J. (2004).
\newblock A breast cancer prediction model incorporating familial and personal
  risk factors.
\newblock {\em Statistics in medicine} {\bf 23,} 1111--1130.

\bibitem[\protect\citeauthoryear{Waldron, Riester, Ramos, Parmigiani, and
  Birrer}{Waldron et~al.}{2016}]{waldron2016doppelganger}
Waldron, L., Riester, M., Ramos, M., Parmigiani, G., and Birrer, M. (2016).
\newblock The doppelg{\"a}nger effect: hidden duplicates in databases of
  transcriptome profiles.
\newblock {\em JNCI: Journal of the National Cancer Institute} {\bf 108,}.

\bibitem[\protect\citeauthoryear{Williams-Blangero and
  Blangero}{Williams-Blangero and Blangero}{2006}]{williams2006collection}
Williams-Blangero, S. and Blangero, J. (2006).
\newblock Collection of pedigree data for genetic analysis in isolate
  populations.
\newblock {\em Human Biology} {\bf 78,} 89--101.

\bibitem[\protect\citeauthoryear{Yan, Lee, Kan, and Giles}{Yan
  et~al.}{2007}]{adaptive_sn}
Yan, S., Lee, D., Kan, M.-Y., and Giles, L.~C. (2007).
\newblock Adaptive sorted neighborhood methods for efficient record linkage.
\newblock In {\em Proceedings of the 7th ACM/IEEE-CS joint conference on
  Digital libraries}, pages 185--194. ACM.

\end{thebibliography}
\bibliographystyle{biom}

\newpage

\beginsupplement

\begin{center}
    \Huge{Supplementary Materials}
\end{center}

\begin{figure}
    \centering
    \includegraphics[width=0.8\textwidth]{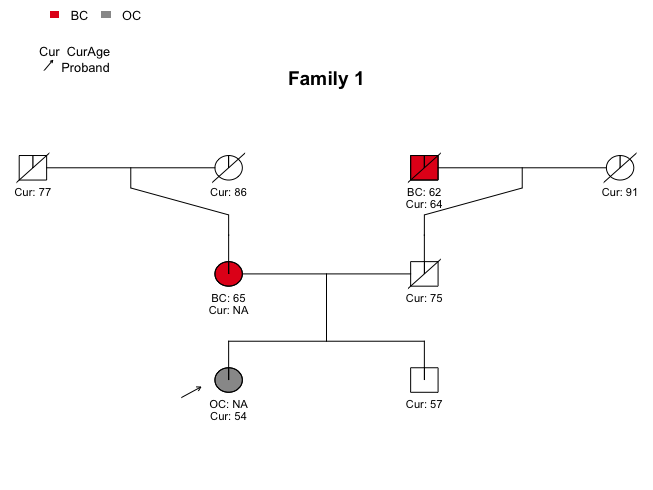}
    \includegraphics[width=0.8\textwidth]{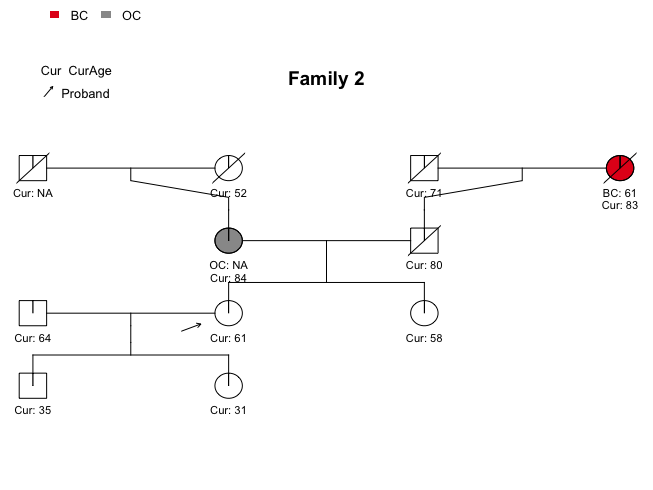}
    \caption{Plots of the example pedigrees in Table~\ref{tab:example}.}
    \label{sfig:example_plot}
\end{figure}

\begin{figure}
    \centering
    \includegraphics[width=\textwidth]{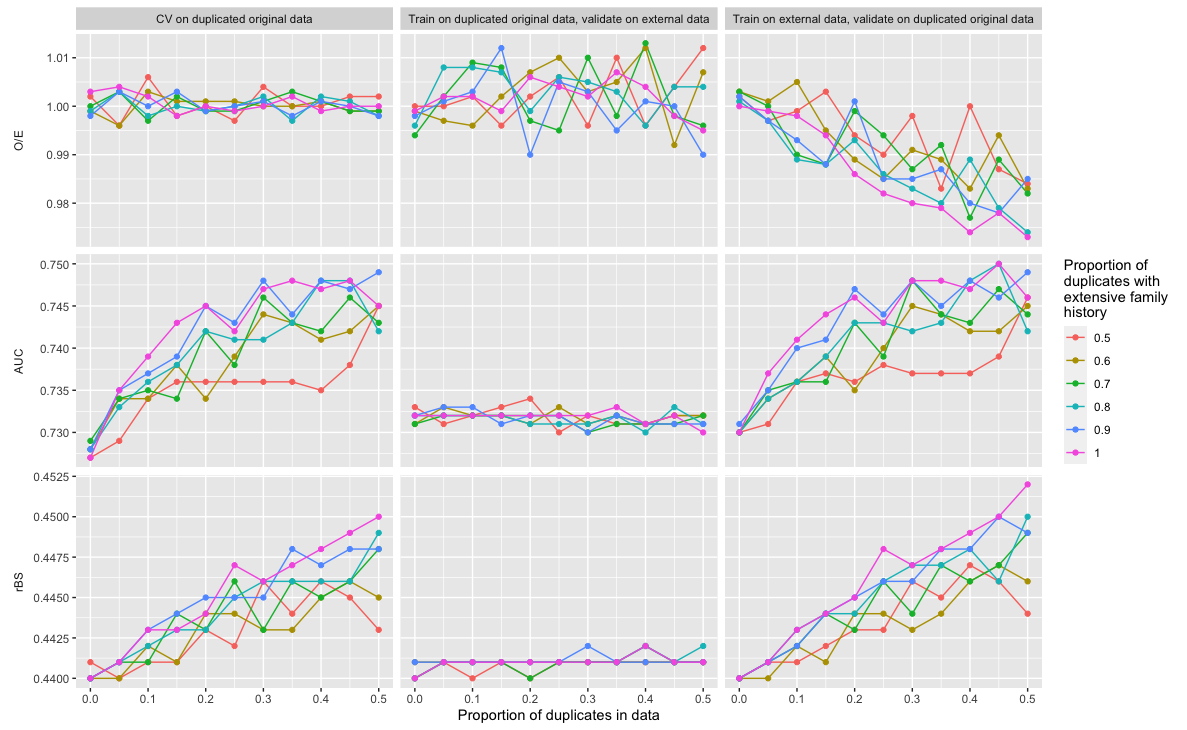}
    \caption{Performance metrics for the logistic regression model under three training/validation scenarios. The duplicated original data is augmented with duplicates, and the external data does not contain duplicates. Carrier statuses are generated with \textit{BRCA1} and \textit{BRCA2} allele frequencies of 0.1. Extenstive family history is defined as having at least 90th percentile of the proportion of family members with breast or ovarian cancer. See Section~\ref{subsec:motiv_results} for the definitions of the three metrics.}
    \label{sfig:sim_effect_af01}
\end{figure}

\begin{table}[ht]
\centering
\begin{tabulary}{\linewidth}{RRRRRRRRRRRR}
  \toprule
& & & & & \multicolumn{3}{c}{Pairwise} & \multicolumn{3}{c}{Cluster} & \\
\cmidrule(lr){6-8} \cmidrule(lr){9-11}
Sliding Window Size & Number of Iterations & Key Length & Threshold & Runtime & Precision & Recall & $F_1$ & Precision & Recall & $F_1$ & GMD \\
  \midrule
20 & 1 & 29 & 6 & 191.50 & 1.000 & 0.998 & 0.999 & 0.999 & 1.000 & 0.999 & 2.4 \\ 
  20 & 1 & 10 & 6 & 193.45 & 1.000 & 0.998 & 0.999 & 0.999 & 0.999 & 0.999 & 3.0 \\ 
  15 & 1 & 20 & 6 & 151.26 & 1.000 & 0.997 & 0.999 & 0.999 & 0.999 & 0.999 & 3.2 \\ 
  20 & 1 & 20 & 6 & 211.35 & 1.000 & 0.998 & 0.999 & 0.999 & 0.999 & 0.999 & 3.4 \\ 
  15 & 1 & 10 & 6 & 159.55 & 1.000 & 0.996 & 0.998 & 0.998 & 0.999 & 0.999 & 4.0 \\ 
  15 & 1 & 29 & 6 & 155.33 & 1.000 & 0.996 & 0.998 & 0.998 & 0.999 & 0.999 & 4.8 \\ 
  20 & 1 & 29 & 5 & 191.50 & 0.999 & 1.000 & 1.000 & 1.000 & 0.999 & 0.999 & 4.8 \\ 
  15 & 1 & 20 & 5 & 151.26 & 0.998 & 1.000 & 0.999 & 0.999 & 0.998 & 0.999 & 9.0 \\ 
  15 & 1 & 10 & 5 & 159.55 & 0.996 & 1.000 & 0.998 & 0.999 & 0.998 & 0.998 & 12.4 \\ 
  10 & 1 & 20 & 5 & 113.85 & 1.000 & 0.982 & 0.991 & 0.994 & 0.997 & 0.995 & 16.8 \\ 
   \bottomrule
   \end{tabulary}
   \caption{Table of the 10 parameter configurations with the lowest GMD values in the simulation in Section~\ref{sec:simulations}.}
\label{stab:sim_params}
\end{table}

\begin{figure}
    \centering
    \includegraphics[width=\textwidth]{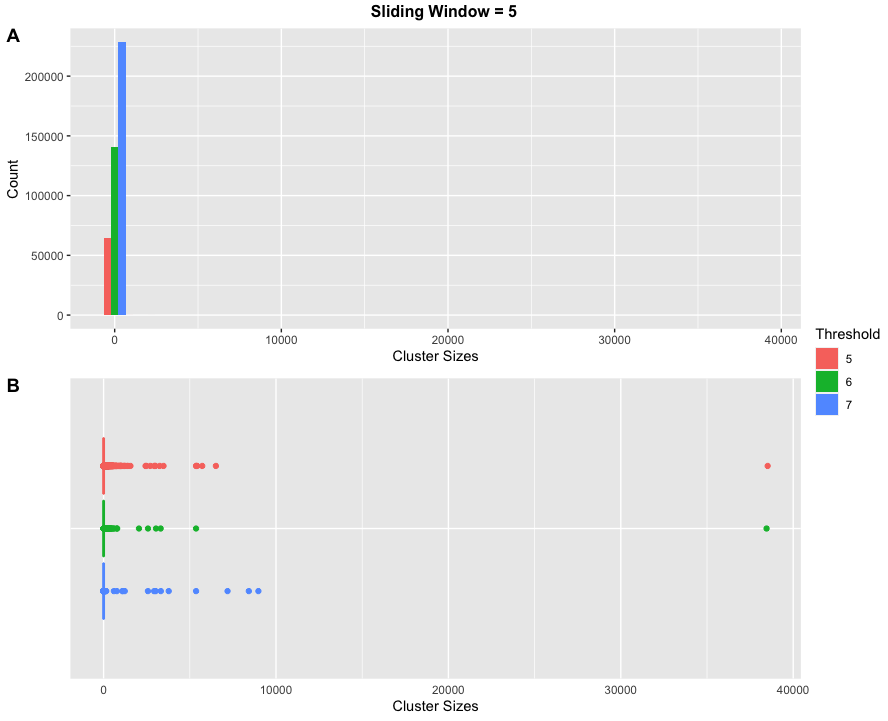}
    \caption{Cluster sizes and counts for the Risk Service Data, with a sliding window size of 5 and thresholds of 5, 6, and 7. Subfigure (A) provides a histogram, with cluster sizes $>1$ invisible because of the skewness of the distribution. Subfigure (B) provides a boxplot that shows the large cluster sizes.}
    \label{sfig:cluster_5}
\end{figure}

\begin{figure}
    \centering
    \includegraphics[width=\textwidth]{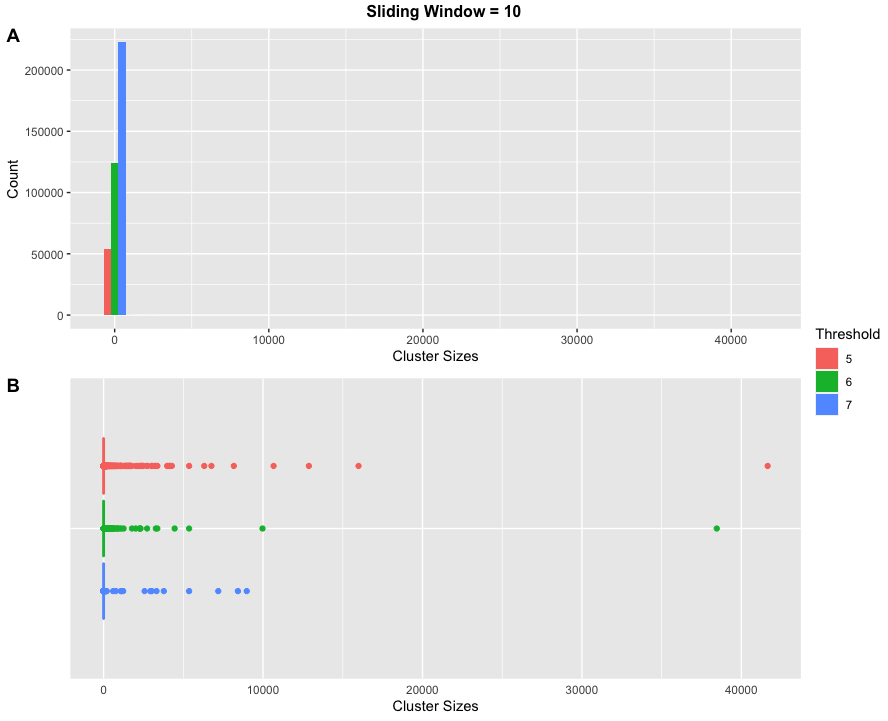}
    \caption{Cluster sizes and counts for the Risk Service Data, with a sliding window size of 10 and thresholds of 5, 6, and 7. Subfigure (A) provides a histogram, with cluster sizes $>1$ invisible because of the skewness of the distribution. Subfigure (B) provides a boxplot that shows the large cluster sizes.}
    \label{sfig:cluster_10}
\end{figure}

\begin{table}[ht]
\centering
\textbf{Sliding window size = 20}
\begin{tabulary}{\linewidth}{RRRRRRRRRRRR}
  \toprule
  & & & & \multicolumn{4}{c}{Proband} & \multicolumn{4}{c}{Relatives} \\
  \cmidrule(lr){5-8} \cmidrule(lr){9-12}
Cluster Size & Time Difference & Average Family Size & Same Family Size & Same Cancer Statuses & Same Ages & Any Cancer & Any Known Ages & Same Cancer Statuses & Same Ages & Any Cancer & Any Known Ages \\
  \midrule
8978 & 21.38 & 7 & Yes & Yes & Yes & No & Yes & Yes & Yes & No & No \\ 
  8424 & 21.47 & 7 & Yes & Yes & Yes & No & Yes & Yes & Yes & No & No \\ 
  7190 & 15.31 & 7 & Yes & Yes & Yes & No & Yes & Yes & Yes & No & No \\ 
  5363 & 21.22 & 20 & Yes & Yes & Yes & No & Yes & Yes & Yes & Yes & Yes \\ 
  3778 & 19.23 & 7 & Yes & Yes & Yes & No & Yes & Yes & Yes & No & No \\ 
  3320 & 16.57 & 7 & Yes & Yes & Yes & No & Yes & Yes & Yes & No & No \\ 
  3021 & 21.90 & 25 & Yes & Yes & Yes & No & Yes & Yes & Yes & No & No \\ 
  2931 & 16.38 & 7 & Yes & Yes & Yes & No & Yes & Yes & Yes & No & No \\ 
  2574 & 7.03 & 15 & Yes & Yes & Yes & No & Yes & Yes & Yes & No & No \\ 
  1228 & 15.99 & 7 & Yes & Yes & Yes & No & Yes & Yes & Yes & No & No \\ 
  1097 & 15.67 & 7 & Yes & Yes & Yes & No & Yes & Yes & Yes & No & No \\ 
  1097 & 15.64 & 7 & Yes & Yes & Yes & No & Yes & Yes & Yes & No & No \\ 
  769 & 2.37 & 7 & Yes & Yes & Yes & Yes & Yes & Yes & Yes & No & No \\ 
  596 & 2.13 & 7 & Yes & Yes & Yes & No & Yes & Yes & Yes & No & No \\ 
  277 & 1.97 & 7 & Yes & Yes & Yes & No & Yes & Yes & Yes & Yes & Yes \\ 
  157 & 59.53 & 7 & Yes & Yes & Yes & No & Yes & Yes & Yes & No & No \\ 
  146 & 6.13 & 7 & Yes & Yes & Yes & No & Yes & Yes & Yes & No & No \\ 
  143 & 2.77 & 7 & Yes & Yes & Yes & No & Yes & Yes & Yes & No & No \\ 
  100 & 1.75 & 7 & Yes & Yes & Yes & No & Yes & Yes & Yes & No & No \\
   \bottomrule
\end{tabulary}
\caption{Summary of differences within the large clusters (size $\geq 100$) when running SNIP on the Risk Service data, with a threshold of 7 and a sliding window size of 20. Cluster Size is the number of families in the cluster. Time Difference is the number of hours between the first and last entries within each cluster. Average Family Size is the average number of family members for the families in the cluster. Same Family Size is an indicator of whether all the families in the cluster have the same number of family members. Same Cancer Statuses is an indicator of whether all the probands in the cluster have the same cancer statuses (cancer or no cancer), or whether each of the other 6 core relative types in the cluster all have the same cancer statuses. Thus, if in a cluster all of the mothers have cancer and none of the other core relative types have cancer, then Same Cancer Statuses is ``Yes" for both probands and for relatives. Similarly, Same Ages is an indicator of either the probands in the cluster or each of the 6 core relative types in the cluster all having the same current ages. Any Cancer is an indicator of whether any of the probands/core relatives in the cluster have cancer. Any Known Ages is an indicator of whether any of the probands/core relatives in the cluster have a known current age.}
\label{stab:summ_within_cluster_20}
\end{table}

\begin{table}[ht]
\centering
\textbf{Sliding window size = 10}
\begin{tabulary}{\linewidth}{RRRRRRRRRRRR}
  \toprule
  & & & & \multicolumn{4}{c}{Proband} & \multicolumn{4}{c}{Relatives} \\
  \cmidrule(lr){5-8} \cmidrule(lr){9-12}
Cluster Size & Time Difference & Average Family Size & Same Family Size & Same Cancer Statuses & Same Ages & Any Cancer & Any Known Ages & Same Cancer Statuses & Same Ages & Any Cancer & Any Known Ages \\
\midrule
8978 & 21.38 & 7 & Yes & Yes & Yes & No & Yes & Yes & Yes & No & No \\ 
  8424 & 21.47 & 7 & Yes & Yes & Yes & No & Yes & Yes & Yes & No & No \\ 
  7190 & 15.31 & 7 & Yes & Yes & Yes & No & Yes & Yes & Yes & No & No \\ 
  5363 & 21.22 & 20 & Yes & Yes & Yes & No & Yes & Yes & Yes & Yes & Yes \\ 
  3778 & 19.23 & 7 & Yes & Yes & Yes & No & Yes & Yes & Yes & No & No \\ 
  3320 & 16.57 & 7 & Yes & Yes & Yes & No & Yes & Yes & Yes & No & No \\ 
  3021 & 21.90 & 25 & Yes & Yes & Yes & No & Yes & Yes & Yes & No & No \\ 
  2931 & 16.38 & 7 & Yes & Yes & Yes & No & Yes & Yes & Yes & No & No \\ 
  2572 & 18.32 & 15 & Yes & Yes & Yes & No & Yes & Yes & Yes & No & No \\ 
  1228 & 15.99 & 7 & Yes & Yes & Yes & No & Yes & Yes & Yes & No & No \\ 
  1097 & 15.67 & 7 & Yes & Yes & Yes & No & Yes & Yes & Yes & No & No \\ 
  1097 & 15.64 & 7 & Yes & Yes & Yes & No & Yes & Yes & Yes & No & No \\ 
  769 & 2.37 & 7 & Yes & Yes & Yes & Yes & Yes & Yes & Yes & No & No \\ 
  596 & 2.13 & 7 & Yes & Yes & Yes & No & Yes & Yes & Yes & No & No \\ 
  199 & 1.15 & 7 & Yes & Yes & Yes & No & Yes & Yes & Yes & Yes & Yes \\ 
  157 & 59.53 & 7 & Yes & Yes & Yes & No & Yes & Yes & Yes & No & No \\ 
  146 & 6.13 & 7 & Yes & Yes & Yes & No & Yes & Yes & Yes & No & No \\ 
  143 & 2.77 & 7 & Yes & Yes & Yes & No & Yes & Yes & Yes & No & No \\ 
  100 & 1.75 & 7 & Yes & Yes & Yes & No & Yes & Yes & Yes & No & No \\
   \bottomrule
\end{tabulary}
\caption{Summary of differences within the large clusters (size $\geq 100$) when running SNIP on the Risk Service data, with a threshold of 7 and a sliding window size of 10. See the caption for Supplementary Table~\ref{stab:summ_within_cluster_20} for a description of the columns.}
\label{stab:summ_within_cluster_10}
\end{table}

\begin{table}[ht]
\centering
\textbf{Sliding window size = 5}
\begin{tabulary}{\linewidth}{RRRRRRRRRRRR}
  \toprule
  & & & & \multicolumn{4}{c}{Proband} & \multicolumn{4}{c}{Relatives} \\
  \cmidrule(lr){5-8} \cmidrule(lr){9-12}
Cluster Size & Time Difference & Average Family Size & Same Family Size & Same Cancer Statuses & Same Ages & Any Cancer & Any Known Ages & Same Cancer Statuses & Same Ages & Any Cancer & Any Known Ages \\
  \midrule
8978 & 21.38 & 7 & Yes & Yes & Yes & No & Yes & Yes & Yes & No & No \\ 
  8424 & 21.47 & 7 & Yes & Yes & Yes & No & Yes & Yes & Yes & No & No \\ 
  7190 & 15.31 & 7 & Yes & Yes & Yes & No & Yes & Yes & Yes & No & No \\ 
  5363 & 21.22 & 20 & Yes & Yes & Yes & No & Yes & Yes & Yes & Yes & Yes \\ 
  3778 & 19.23 & 7 & Yes & Yes & Yes & No & Yes & Yes & Yes & No & No \\ 
  3320 & 16.57 & 7 & Yes & Yes & Yes & No & Yes & Yes & Yes & No & No \\ 
  3021 & 21.90 & 25 & Yes & Yes & Yes & No & Yes & Yes & Yes & No & No \\ 
  2931 & 16.38 & 7 & Yes & Yes & Yes & No & Yes & Yes & Yes & No & No \\ 
  2572 & 18.32 & 15 & Yes & Yes & Yes & No & Yes & Yes & Yes & No & No \\ 
  1228 & 15.99 & 7 & Yes & Yes & Yes & No & Yes & Yes & Yes & No & No \\ 
  1097 & 15.67 & 7 & Yes & Yes & Yes & No & Yes & Yes & Yes & No & No \\ 
  1097 & 15.64 & 7 & Yes & Yes & Yes & No & Yes & Yes & Yes & No & No \\ 
  769 & 2.37 & 7 & Yes & Yes & Yes & Yes & Yes & Yes & Yes & No & No \\ 
  596 & 2.13 & 7 & Yes & Yes & Yes & No & Yes & Yes & Yes & No & No \\ 
  145 & 4.17 & 7 & Yes & Yes & Yes & No & Yes & Yes & Yes & No & No \\ 
  143 & 2.77 & 7 & Yes & Yes & Yes & No & Yes & Yes & Yes & No & No \\ 
  100 & 1.75 & 7 & Yes & Yes & Yes & No & Yes & Yes & Yes & No & No \\ 
   \bottomrule
\end{tabulary}
\caption{Summary of differences within the large clusters (size $\geq 100$) when running SNIP on the Risk Service data, with a threshold of 7 and a sliding window size of 5. See the caption for Supplementary Table~\ref{stab:summ_within_cluster_20} for a description of the columns.}
\label{stab:summ_within_cluster_5}
\end{table}

\newpage

\section{Greedy Match Score} \label{ssec:greedy}

\subsection{Description} \label{ssubsec:greedy_description}

Since the intersection score in the decision step of the SNIP algorithm only considers the 7 core relative types, we introduce an alternative, which we call the \textit{greedy match score}. The metric leverages the inherent hierarchical nature of the pedigrees in a greedy fashion to quantify potential duplicates, with pairs attaining scores below a user-defined percentile discarded. An example is provided in Supplementary Figure~\ref{sfig:greedy}. We begin by calculating the similarity between the reported probands as the sum of variables containing the same value between the two records. For binary variables, a score of 1 is assigned if the values match while a score of 0 is assigned if they do not match. Continuous variables receive a score of 1 if they match and a score of 1 divided by the difference between the values if they do not match. A weighted sum of the variable match scores is then computed, with the weights derived from the in-sample standard deviation of that variable. This is chosen such that variables with minimal-to-no standard deviation do not highly impact the overall score, although the user may choose any weighting scheme. The score is normalized by the number of variables considered and summed across all reported individuals within the reference pedigree. Next we greedily find the maximal match between the candidate pedigrees by repeating this member-specific scoring procedure for each member of the reference pedigree. Mother and father IDs are used to classify members into the following categories: mother, father, maternal grandmother, maternal grandfather, paternal grandmother, paternal grandfather, children, maternal aunts and uncles, paternal aunts and uncles, cousins, and other. Within a category, all possible comparisons are made between the reference members and the comparison members, and the observations forming the pair which attained the maximum score are subsequently removed from the future scoring comparisons. In this greedy fashion, we iterate through the reference family members and for each family member find the maximum matching family member from the comparison pedigree. All reference family members in a given category without a corresponding comparison match are re-labeled to the ``other" category and greedily scored. The sum over all categories is then computed and used to summarize the likelihood of the potential duplicate pair representing a true duplicate. Lastly, all scores are sorted and scores below a user-supplied percentile threshold are discarded.

\subsection{Simulation} \label{ssubsec:greedy_sim}

We applied the greedy match score as an alternative to the intersection score for the simulations in Section~\ref{sec:simulations}. We used percentile thresholds of 0.75, 0.85, and 0.95. As in Section~\ref{sec:simulations}, we only computed metrics when the algorithm partition had at least half of the number of clusters as the true partition. None of the three percentile thresholds for any of the parameter configurations reached this threshold, and hence none of the metrics were computed. This demonstrates the ineffectiveness of the greedy match score as designed.

\begin{figure}
    \centering
    \includegraphics[width=\textwidth]{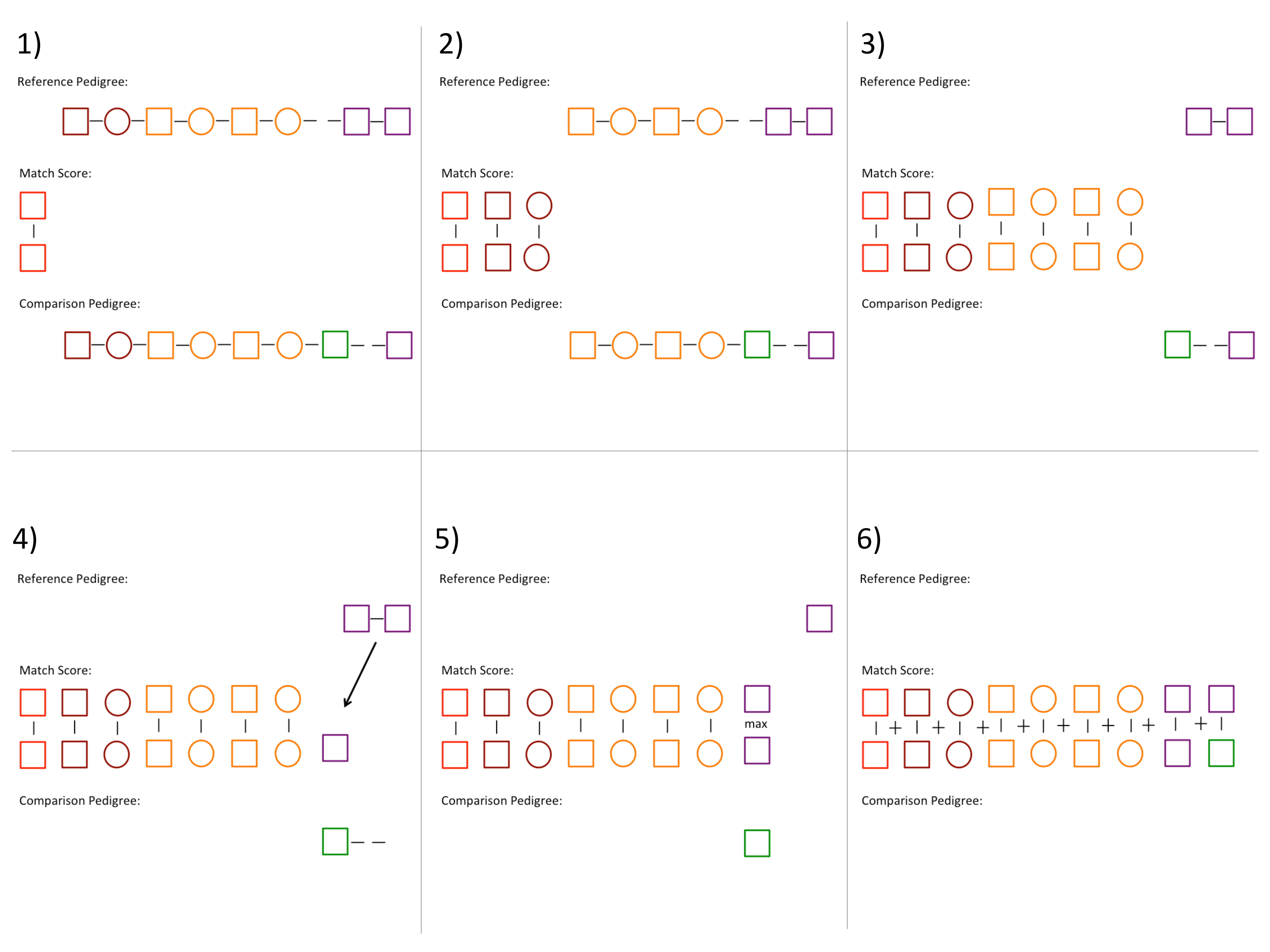}
    \caption{Greedy pedigree scoring example. Shapes represent reported gender (circle for female, and square for male) and colors represent derived relationship type (red = proband, maroon = parental kinships, orange = second-generation kinships, purple = aunts/uncles, green = cousins). Scoring proceeds by finding, for each relationship type in the reference pedigree, the highest-scoring match in the comparison pedigree. Remaining individuals are matched across relationship type (green matched with purple) once all matches within relationship type have been found.}
    \label{sfig:greedy}
\end{figure}

\end{document}